\documentclass[aps,twocolumn,showpacs,groupedaddress, nofootinbib]{revtex4}  
\usepackage{graphicx} 
\usepackage{dcolumn}   
\usepackage{bm}        
\usepackage{amssymb}   
\usepackage{acronym}   
\usepackage{float}
\usepackage{xcolor}
\usepackage{latexsym}
\usepackage{multirow}
\usepackage[flushleft]{threeparttable}
\usepackage{natbib}
\usepackage{subfigure}
\usepackage{amsmath}

\begin{document}

\widetext


\title{Detection and Classification of Supernova Gravitational Waves Signals: A Deep Learning Approach}

\author{Man Leong Chan$^{1}$, Ik Siong Heng$^{2}$ \& Chris Messenger$^{2}$ }
\affiliation{
$^1$Department of Applied Physics, Fukuoka University, Nanakuma 8-19-1, Fukuoka 814-0180, Japan\\
$^2$SUPA, School of Physics and Astronomy, University of Glasgow, Glasgow G12 8QQ, UK}

\begin{abstract}
We demonstrate the application of a convolutional neural network to the
gravitational wave signals from core collapse supernovae. Using simulated time
series of gravitational wave detectors, we show that based on the explosion
mechanisms, a convolutional neural network can be used to detect and classify
the gravitational wave signals buried in noise. For the waveforms used in the
training of the convolutional neural network, our results suggest that a
network of advanced LIGO, advanced VIRGO and KAGRA, or a network of LIGO A+,
advanced VIRGO and KAGRA is likely to detect a magnetorotational core collapse
supernovae within the Large and Small Magellanic Clouds, or a Galactic event if
the explosion mechanism is the neutrino-driven mechanism.  By testing the
convolutional neural network with waveforms not used for training, we show that
the true alarm probabilities are $52\%$ and $83\%$ at $60$ kpc for waveforms
$\text{R3E1AC}$ and $\text{R4E1FC\_L}$.  For waveforms $\text{s}20$ and
$\text{SFHx}$ at $10$ kpc, the true alarm probabilities are $70\%$ and $93\%$
respectively. All at false alarm probability equal to $10\%$. 
\end{abstract}

\pacs{}
\maketitle
\acrodef{GW}[GW]{gravitational wave}
\acrodef{BNS}[BNS]{binary neutron star}
\acrodef{BBH}[BBH]{binary black hole}
\acrodef{NSBH}[NSBH]{neutron star black hole}
\acrodef{EM}[EM]{electromagnetic}
\acrodef{CBC}[CBC]{compact binary coalescence}
\acrodef{CNN}[CNN]{Convolutional neural network}
\acrodef{CCSN}[CCSN]{core collapse supernova}
\acrodefplural{CCSN}[CCSNe]{core collapse supernovae}
\acrodef{ROC}[ROC]{receiver operator characteristic}
\acrodef{TAP}[TAP]{true alarm probability}
\acrodef{FAP}[FAP]{false alarm probability}
\acrodef{aLIGO}[aLIGO]{advanced LIGO}
\acrodef{AdVirgo}[AdVirgo]{advanced VIRGO}

\section{Introduction}
%
%
Since 2015 when LIGO made the first direct observation of \acp{GW} from the
merger of a binary black hole~\cite{abbott2016observation}, there have been
numerous observations of \acp{GW} from similar systems in the subsequent
observation runs of LIGO and VIRGO~\cite{abbott2016gw151226,
abbott2017gw170608, abbott2017gw170814}. These discoveries represent a crucial
milestone in \ac{GW} astronomy and have opened up a new window on the universe.
More recently, LIGO and VIRGO have observed the \acp{GW} from a binary neutron
star merger~\cite{abbott2017gw170817, abbott2017gravitational, abbott2017multi}
where the \acp{GW} and an associated gamma-ray burst were observed
simultaneously. Other counterparts across the electromagnetic spectrum were
also observed by later follow-up observations~\cite{abbott2017multi}. In the
near future, many more observations of \acp{GW} from similar \ac{CBC} systems
can be expected as KAGRA starts joint observations with LIGO and
VIRGO~\cite{aso2013interferometer, somiya2012detector, abbott2018prospects}.

%
%
In addition to \acp{CBC}, massive stars with $10-100 \text{M}_\odot $ at
zeros-age main sequence ending their lives by becoming 
\acp{CCSN} are also considered to be potential sources to the second generation
detectors such as \ac{aLIGO}~\cite{aasi2015advanced},
\ac{AdVirgo}~\cite{acernese2014advanced} and KAGRA
interferometers~\cite{aso2013interferometer, gossan2016observing,
abbott2016first, abbott2019optically}. It is currently not entirely clear to astronomers how such
massive stars become supernovae. The basic theory of the explosion, confirmed
by the neutrino events observed from SN1987A~\cite{sato1987analysis}, begins
with a massive star at the final stage of its life forming a core that is
composed of iron nuclei after it has burned all its stellar fuel via fusion
reactions. The iron core is supported by the pressure of relativistic
degenerate electrons and if the mass of the core exceeds the effective
Chandrasekhar mass~\cite{baron1990effect, bethe1990supernova}, core collapse
will ensue and continue until the core reaches nuclear density. The nuclear
equation of state will then stiffens by the strong nuclear force above the
nuclear density and stops the core collapse. The inner core will bounce back
and a shock wave will be sent through the infalling matter.  By losing energy
to the dissociation of the iron nuclei and to neutrino cooling, the shock wave
will stall. For the star to become a supernova, the shock wave will need to be
revived~\cite{o2011black}. However the mechanism via which this occurs and the
supernova explosion is caused has been the subject of intense study and remains
an unsolved problem.

%
%
There exist two most popular theories for how the shock is revived and a star becomes supernova, 
the neutrino-driven mechanism~\cite{bethe1985revival, bethe1990supernova} and the
magnetorotational mechanism~\cite{janka2012explosion, kotake2012core,
mezzacappa2014two}. For supernova progenitors with core rotation too slow to
affect the dynamics~\cite{takiwaki2016three, summa2018rotation}, the
neutrino-driven mechanism is believed to be the active mechanism. The majority
of the observed \acp{CCSN} can be explained by the neutrino
mechanism~\cite{bruenn2016development}. The neutrino
mechanism~\cite{bethe1985revival, janka2007theory} suggests that approximately
$5\%$ of the outgoing neutrino luminosity is stored below the shock, which
causes turbulence to occur and thermal pressure to increase. The stalled shock
can be revived by their combined effects~\cite{couch2015role}. Producing a
\ac{CCSN} via the neutrino mechanism may also require convection and the
standing accretion shock instability~\cite{blondin2003stability}. On the other
hand, the magnetorotational mechanism requires rapid core spin and a strong
magnetic field~\cite{leblanc1970numerical, burrows2007simulations,
takiwaki2009special,
moiseenko2006magnetorotational,mosta2014magnetorotational}. Together, they may
produce an outflow that may cause some of the most energetic \acp{CCSN}
observed and may be able
to explain the extreme hypernovae and the observed long gamma-ray
bursts~\cite{woosley2006progenitor, yoon2005evolution, de2013rotation}.
 
%
%
Correctly classifying the \ac{GW} signals from a \ac{CCSN} is important in
understanding the explosion mechanism.
As \acp{GW} are generated in the central core of a \ac{CCSN}, they
are likely to carry direct information of the \ac{CCSN} and therefore provide a
probe of the explosion mechanism that produces them. In \ac{GW} astronomy, for
\ac{CBC} events the established method for searching for signals is
matched-filtering~\cite{usman2016pycbc, cannon2012toward}. However, since the
emission process of the \acp{GW} from \acp{CCSN} is affected by turbulence in
the post-bounce and is expected to be stochastic in nature,  the signal
evolution cannot be predicted robustly~\cite{ott2009gravitational,
kotake2013multiple, kotake2009stochastic}.  This in turn prevents
matched-filtering from being applied to \acp{CCSN}.
Methods and algorithms have been developed for the detection and classification
of signals from \acp{CCSN}.  For example, a method known as principle component
analysis has been developed~\cite{heng2009rotating, rover2009bayesian,
powell2015classification, powell2017classification,
suvorova2019reconstructing}. This method creates a set of component basis
vectors from a larger set of \ac{CCSN} waveforms belonging to a
particular mechanism where the basis vectors represent the common features of
those waveforms.  There have been other approaches developed in the literature
such as Bayesian inference~\cite{rover2009bayesian, cannon2012toward}, Bayesian model
selection~\cite{logue2012inferring}, multivariate regression
modelling~\cite{engels2014multivariate}, maximum
entropy~\cite{summerscales2008maximum}, maximum
likelihood~\cite{klimenko2016method} and Tikhonov regularization
scheme~\cite{rakhmanov2006rank, hayama2007coherent}.  

%
%
In recent years, the field of machine learning and its sub-field, deep
learning, have been rapidly developing and have shown great potential in many
scientific fields~\cite{krizhevsky2012imagenet, NIPS2014_5423,
simonyan2014very, chen2014semantic, zeiler2014visualizing, szegedy2015going}.
For example, deep learning has been successfully applied to fields including
medical diagnosis~\cite{kononenko2001machine}, object
detection~\cite{redmon2016you}, image
recognition/processing/generation~\cite{he2016deep, krizhevsky2012imagenet,
zhang2016colorful, karpathy2015deep}, and  language
processing~\cite{lample2016neural}. In \ac{GW} astronomy, deep learning has
mostly been applied to the identifications of detector noise artefacts
(glitches)~\cite{mukund2017transient, zevin2017gravity, george2017deep} and the
detection of astrophysical signals~\cite{george2018deep, gabbard2018matching, astone2018new},
and their parameter estimation~\cite{2019arXiv190906296G}.
An advantage of the use of a \ac{CNN} in the detection
of a \ac{GW} signal is that a \ac{CNN} is relatively computationally cheap
compared to other more traditional methods. This is because the heavy
computational work is usually done during the training stage of a \ac{CNN}
prior to its actual application~\cite{goodfellow2016deep}.

%
%
In this work, we demonstrate that a \ac{CNN} can be applied to the detection of
\acp{GW} from \acp{CCSN} and to the classification of their explosion
mechanisms. 
To train our \ac{CNN}, we use simulated \ac{CCSN} waveforms from a
number of studies and simulate sources at a range of distances from $10$ to
$200$ kpc for two networks of four \ac{GW} detectors. The first network
consists of \ac{aLIGO}, \ac{AdVirgo} and KAGRA. For the second network, we
still include the detectors of \ac{AdVirgo} and KAGRA, but replace the two
detectors of \ac{aLIGO} with a modest set of planned upgrade version of them -
LIGO A+ in Hanford and Livingston~\cite{miller2015prospects, LIGOW}. All detectors 
are at their design sensitivities. We then
apply the \ac{CNN} to waveforms excluded from the training set to show 
the performance on an independent testing set.
Similar to~\cite{astone2018new}, 
we use machine learning techniques for the search of \acp{GW} from \acp{CCSN}.
However, the \ac{CNN} developed in ~\cite{astone2018new} takes time-frequency images of \ac{GW} detector 
data from the coherent WaveBurst pipeline~\cite{klimenko2005constraint} as input, 
while we aim to develop a \ac{CNN} that works independently and takes time series data from \ac{GW} detectors as input.

%
%
The remaining of this paper is structured as follows. In Section~\ref{sec:CNN},
we present a brief explanation of the concept of a \ac{CNN} as well as the
\ac{CNN} architecture  used for this study. In Section~\ref{sec:spwf}, we
introduce the waveforms we use for the training of the \ac{CNN} and  also
discuss the procedure with which we generated the input data for the \ac{CNN}.
The results are shown in Sections~\ref{sec:result} and~\ref{sec:unseen},
followed by our conclusions in Section~\ref{sec:conclusion}.

\section{Convolutional Neural network}\label{sec:CNN}

%
%
A \ac{CNN} is a computational processing system that takes in data and 
is able to classify the input data as one of the N types it has learnt through training.
A \ac{CNN} is composed of interconnected layers of computational nodes~\cite{o2015introduction}. 
The nodes are known as neurons and the outputs of which are processed with an
activation function. The activation function performs an elementwise non-linear
operation to the output of the layer. 
There are commonly three types of layers: convolutional layers, max-pooling layers, and
fully connected layers~\cite{o2015introduction}. Convolutional layers perform
the mathematical operation of convolution between the weights of the layer's
neurons and the input to that layer. Max-pooling layers perform a down-sampling
process that reduces the size of the data by selecting the maximum of data
samples within fixed size bins. It can reduce the computational cost by
decreasing the number of trainable parameters (weights and biases) of the
\ac{CNN}.  Fully connected layers are layers that connect every neuron in its
layer to every neuron of its immediate previous and next layer. 

When multiple layers of these three types are stacked and connected one after
the other, a \ac{CNN} has been formed, where the output of each layer is the
input of the next layer. 
%
%
How these layers are connected in a \ac{CNN}, the
numbers of layers, their type, the number of neurons, the convolutional filter
size, max-pooling size, and the activation functions used, describes the
architecture of the \ac{CNN} (also known as hyper-parameters).  
In a standard \ac{CNN}, 
the first layer, also known as the input layer and often a convolutional layer, 
takes in raw values of the input (or time series from \ac{GW} detectors in our case, this will be explained more in section \ref{sec:spwf}).
The output layer is usually a fully connected layer
with a softmax activation function with output used to represent the class
scores or probabilities in the case of detection and classification.
For a \ac{CNN} designed to classify its input into different classes, 
the output layer computes the probabilities of the inferred classes.
The architecture of the remainder of the \ac{CNN} should depend on the specific task that it is being
trained to solve. An over-complicated model with too many trainable parameters
makes it easier to result in over-fitting, where the network
essentially memorises the training data but is unable to generalise to new
data. Alternatively, an overly simple architecture will struggle to capture
the features inherent to the input and not perform well in classification.
Automated schemes do exist for finding the optimal combination of
hyperparameters but in general and in our case the final architecture is
often obtained through rough trial and error, followed by fine tuning.

\begin{figure*}
\includegraphics[width=0.96\textwidth]{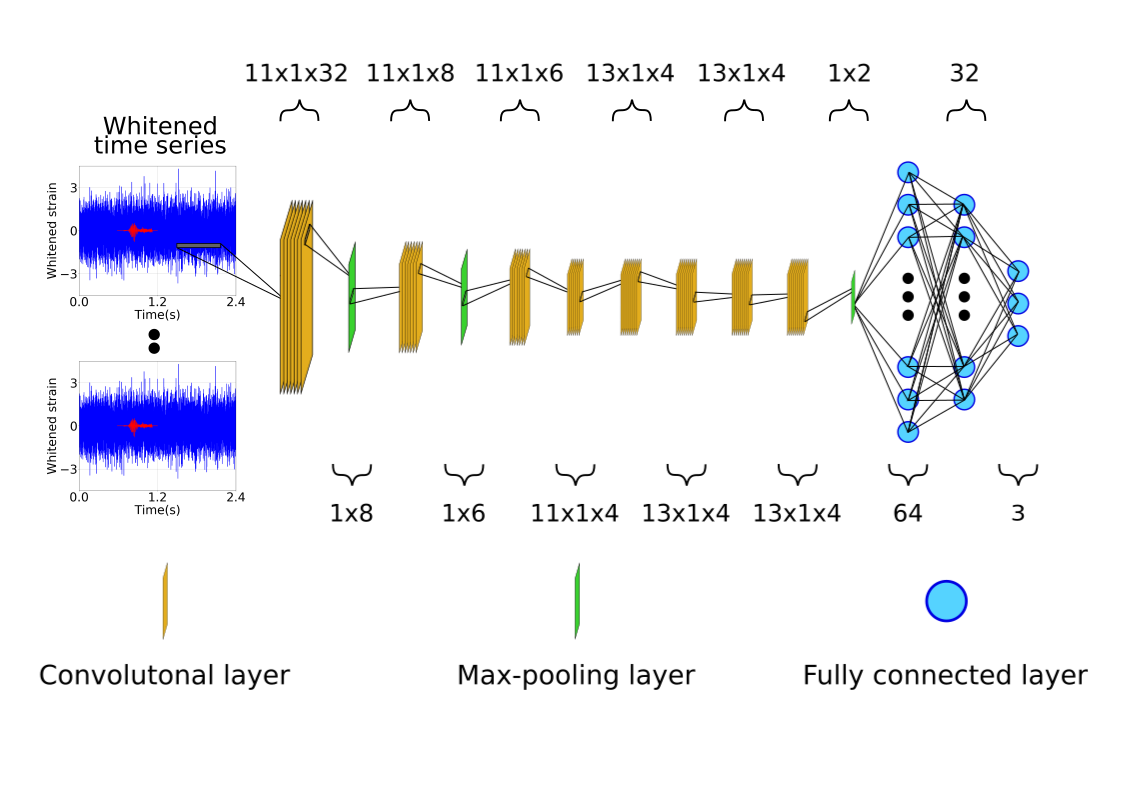}
\caption{An illustration of the architecture of the \ac{CNN} used in this paper
for the detection and classification of \ac{CCSN} \ac{GW} signals in noisy data. The
\ac{CNN} consists of $8$ convolutional layers, $3$ max-pooling layers and $3$
fully connected layer including the output layer.  The input layer takes the
simulated time series of the detectors as input, feeding through the \ac{CNN}.
The \ac{CNN} will output three probabilities at the last layer.  The numbers
above or below each layer indicate the kernal size of the layer. For example,
the first convolutional layer has $11$ filters, each of which is $1$ by $32$ in
size.
\label{fig:CNN}}
\end{figure*}

%
%
During the training stage, the weights of the neurons in a \ac{CNN} are updated
using an algorithm called back-propagation~\cite{lecun1988theoretical}. The
outputs of the \ac{CNN} (the probabilities of the inferred classes in our case) are used as an input to the loss-function that is associated with the entire network.  
The value of the loss function is used to evaluate how well the algorithm models the input data of the \ac{CNN}.
After a subset of training data is fed through the \ac{CNN}, the back-propagation algorithm 
then computes the gradient of the loss-function with respect
to the trainable weights and biases within the network. 
The size of the subset of the training data is referred to as batch.
A gradient descent algorithm is then used to adjust the values of the weights and biases of the
neurons in each layer in order to iteratively minimise the loss-function.  
When the loss function is minimised and training is therefore complete, the \ac{CNN}
will be able to take in input in the form of new data and its output will best
represent the probability of that data belonging to each of the trained
classes. 
A well performing network will give high output probabilities to the
correct class in most cases. The process of achieving the minimisation of the
loss function during the training stage is the process whereby the machine is
\textquotedblleft learning\textquotedblright.
%
%
In this work, we employ a \ac{CNN} of $8$ convolutional layers, $3$ max-pooling
layers, and $3$ fully connected layers. We also use a technique, known as drop-out, for addressing over-fitting~\cite{srivastava2014dropout}. 
This terms refers to the way this technique works - 
by removing units of random choices temporarily in the hidden layers and their associated connections in the \ac{CNN}.
The exact architecture of the \ac{CNN}
is shown in Table~\ref{table:architecture} and illustrated in
Fig.~\ref{fig:CNN}. 
%
%
\begin{table}[]
\centering
\begin{threeparttable}
\caption{The architecture of the \ac{CNN}}
\label{table:architecture}
\begin{tabular}{ccccc}
\\
\toprule
Layer & Type      & Neurons  & Filter size & Act. Fun \\ \hline
1     & Conv      & 11        & 32          & Elu      \\ 
2     & Max-pool  &          & 8          &          \\
3     & Conv      & 11        & 8          & Elu      \\
4     & Max-pool  &          & 6           &          \\
5     & Conv      & 11       & 6           & Elu      \\
6     & Conv      & 11       & 4           & Elu      \\
7     & Conv      & 13       & 4           & Elu      \\
8     & Conv      & 13       & 4           & Elu      \\
9     & Conv      & 13       & 4           & Elu      \\
10    & Conv      & 13       & 4           & Elu      \\
11    & Max-pool  &          & 2           &          \\
12    & Fully-con & 64(50\%) &             & Elu      \\
13    & Fully-con & 32(50\%) &             & Elu      \\
14    & Fully-con & 3        &             & Softmax      \\ 
\hline
\hline
\end{tabular}
\begin{tablenotes}
\setlength\labelsep{0pt}
\normalfont{
\item The architecture of the \ac{CNN} used in this work for the purpose of
distinguishing supernova signals in backgound \ac{GW} detector noise vs detectr
noise alone. In the table, Conv means convolutional layer, Max-pool means
max-pooling layer, and Fully-con means fully-connected layer. Neuron and Act.
Fun indicate the number of neurons and the activation function used for the
layer respectively. The numbers in the bracket for the fully-connected layers
are the number used for drop-out.}
\end{tablenotes}
\end{threeparttable}
\end{table}
%
%

The problem we are trying to solve is a problem of multi-class
classification, and hence the loss function employed for this work is the
categorical cross-entropy~\cite{abadi2016tensorflow}, defined as
\begin{equation}\label{eq:cce}
 L(y,\hat{y}) = -\sum^M_{j=1}\sum^C_{i=1}y_{ij}\log(\hat{y}_{ij}),
\end{equation}
where $C$ is the number of the classes and $M$ is the number of the batch.
For the $j^{\text{th}}$ sample and the $i^{\text{th}}$ class, $y_{ij}$ is the corresponding class value. It is equal to $1$ for the
true class and $0$ otherwise.  Similarly, $\hat{y}_{ij}$ is the predicted
probability from the \ac{CNN} for the $i^{\text{th}}$ class and the $j^{\text{th}}$
sample.

\section{Data}\label{sec:spwf}
%
%
We establish a \ac{CNN} for the purpose of distinguishing detector time series
among three classes, i.e., magnetorotational signals $+$ background noise,
neutrino-driven signals $+$ background noise, and pure background noise. For
this purpose, it is necessary to prepare training, validation and testing data
of these three classes. The training data is used for tuning the weights of
the neurons in the layers in the \ac{CNN}. Validating data is usually applied during training 
to verify that the \ac{CNN} is learning the features inherent to the data and
to prevent the \ac{CNN} from over-fitting the training data.
For a \ac{CNN} that is not over-fitting, the value of the loss function will be close between the validation data and training data.
Testing data is to test the performance of the trained \ac{CNN} and applied after training has completed. 

%
%
In our case, we define an input data sample as a set of simulated time series
stacked together as a $k \times p$ matrix where $k$ is the number of detectors
and $p$ the number of elements in the time series. To this end, we use
simulated waveforms taken from the literature. The magnetorotational \ac{CCSN}
signals are taken from~\cite{abdikamalov2014measuring,
dimmelmeier2008gravitational, richers2017equation}\footnote{\href{https://stellarcollapse.org/ccdiffrot}
{https://stellarcollapse.org/ccdiffrot}}$^{,}$\footnote{\href{https://zenodo.org/record/201145}{https://zenodo.org/record/201145}}. 
The simulations
in~\cite{abdikamalov2014measuring} were focused on the dependence of the
waveforms on the angular momentum distribution of the progenitors and provide
us with 92 waveforms. In~\cite{dimmelmeier2008gravitational}, 136 waveforms are
available from investigations into a variety of rotation rates and masses of
the progenitors. The simulations in~\cite{richers2017equation} covered a
parameter space of $18$ different equations of state and $98$ rotation profiles
for a progenitor of $12\text{M}_\odot$ providing a total of $1824$ waveforms. All
the simulations from the studies were based on 2D simulations of supernovae.
It needs to be pointed out that as the authors in these studies
were interested in the early stage of post-core bounce, and/or the effects of the equation of states on the \acp{GW} generated at core bounce,
the simulations were only run for a short amount of time after core bounce. 
For example, the simulations in~\cite{richers2017equation} were only run for $50$ ms, and only approximately up to $10$ ms after core bounce were used.

\begin{figure*}
     \begin{center}
        \subfigure[]{
            \label{fig:magsample1}
            \includegraphics[width=0.315\textwidth]{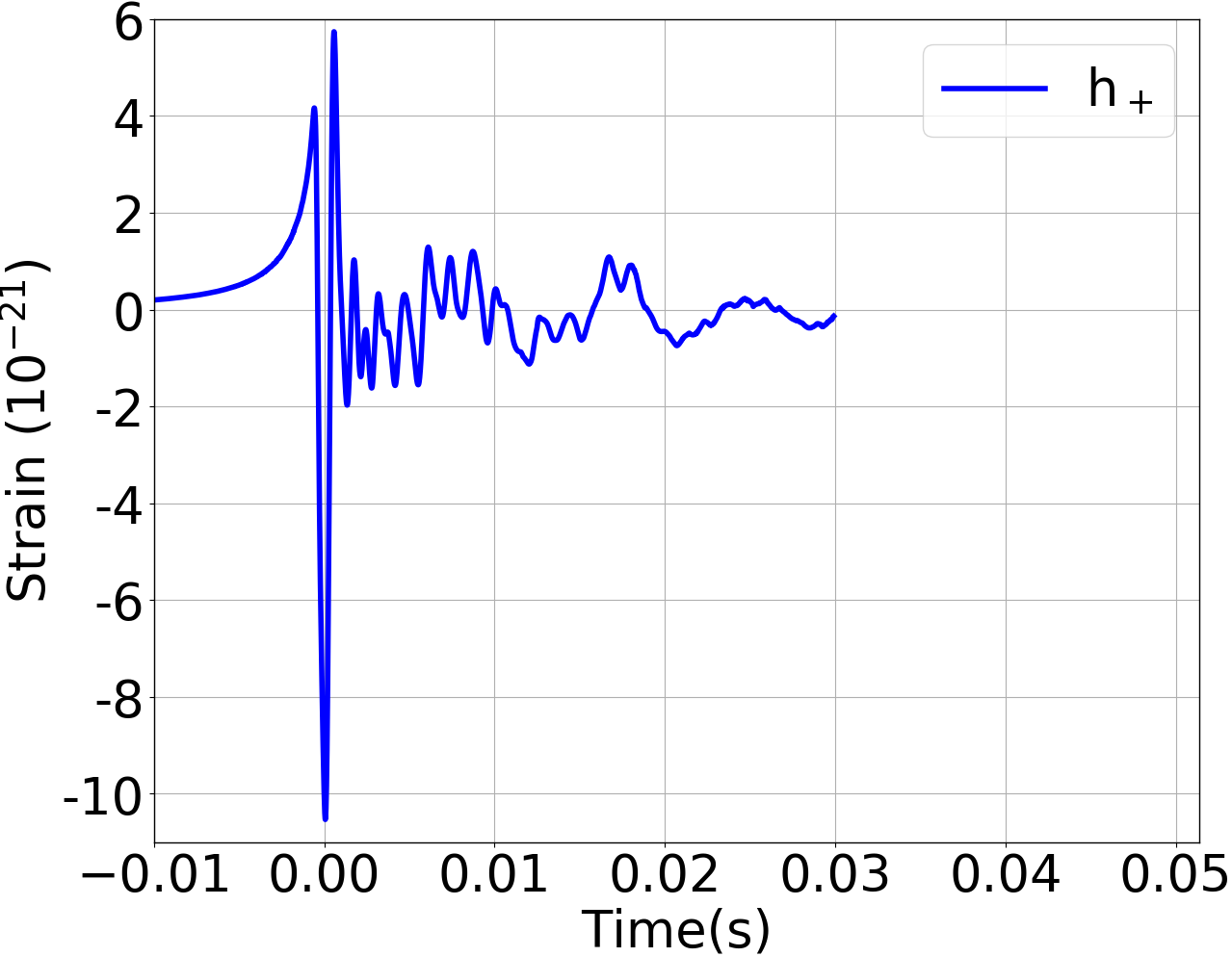}%
        }\quad
        \subfigure[]{
            \label{fig:magsample2}
            \includegraphics[width=0.315\textwidth]{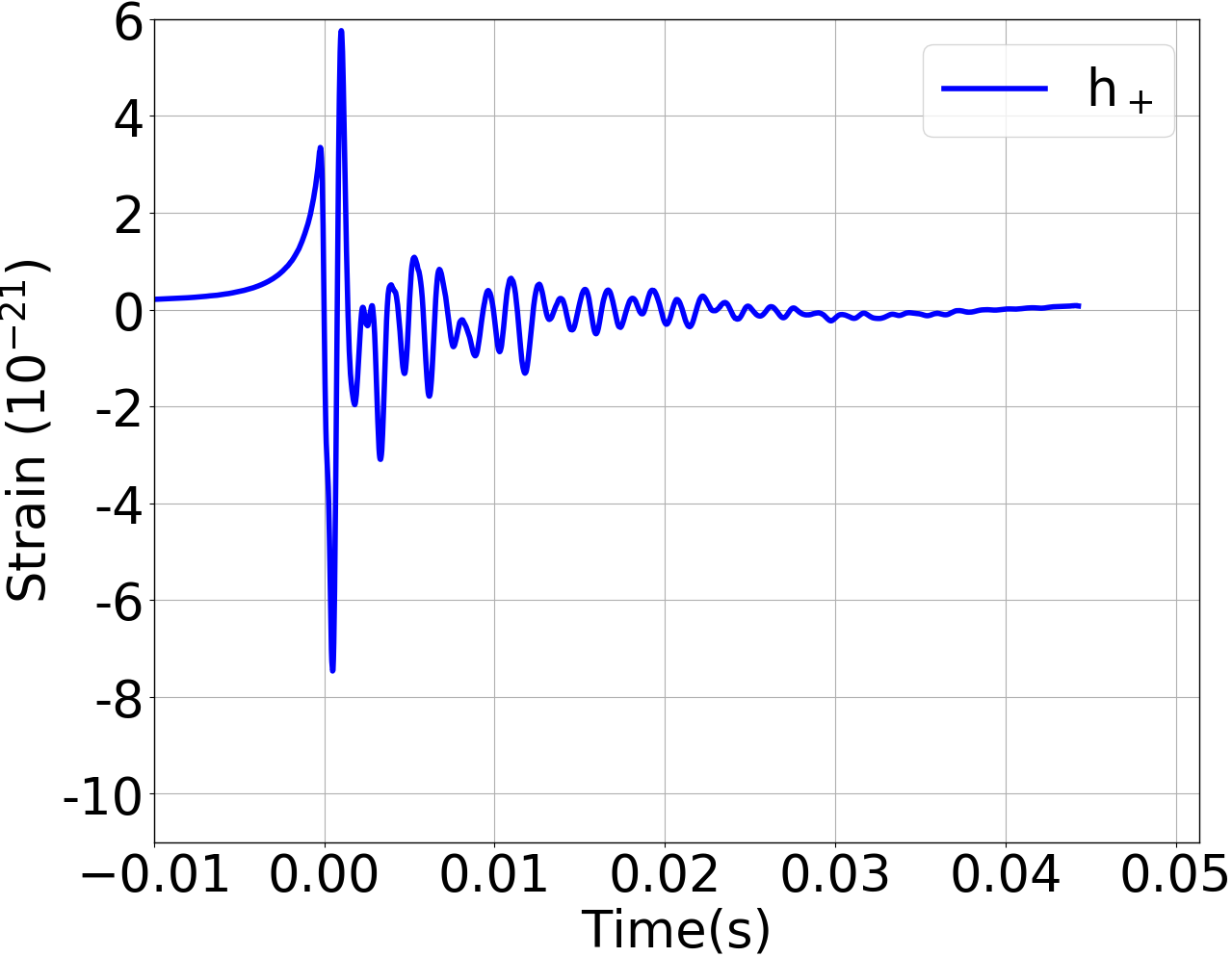}%
        }\quad
        \subfigure[]{
            \label{fig:magsample3}
            \includegraphics[width=0.315\textwidth]{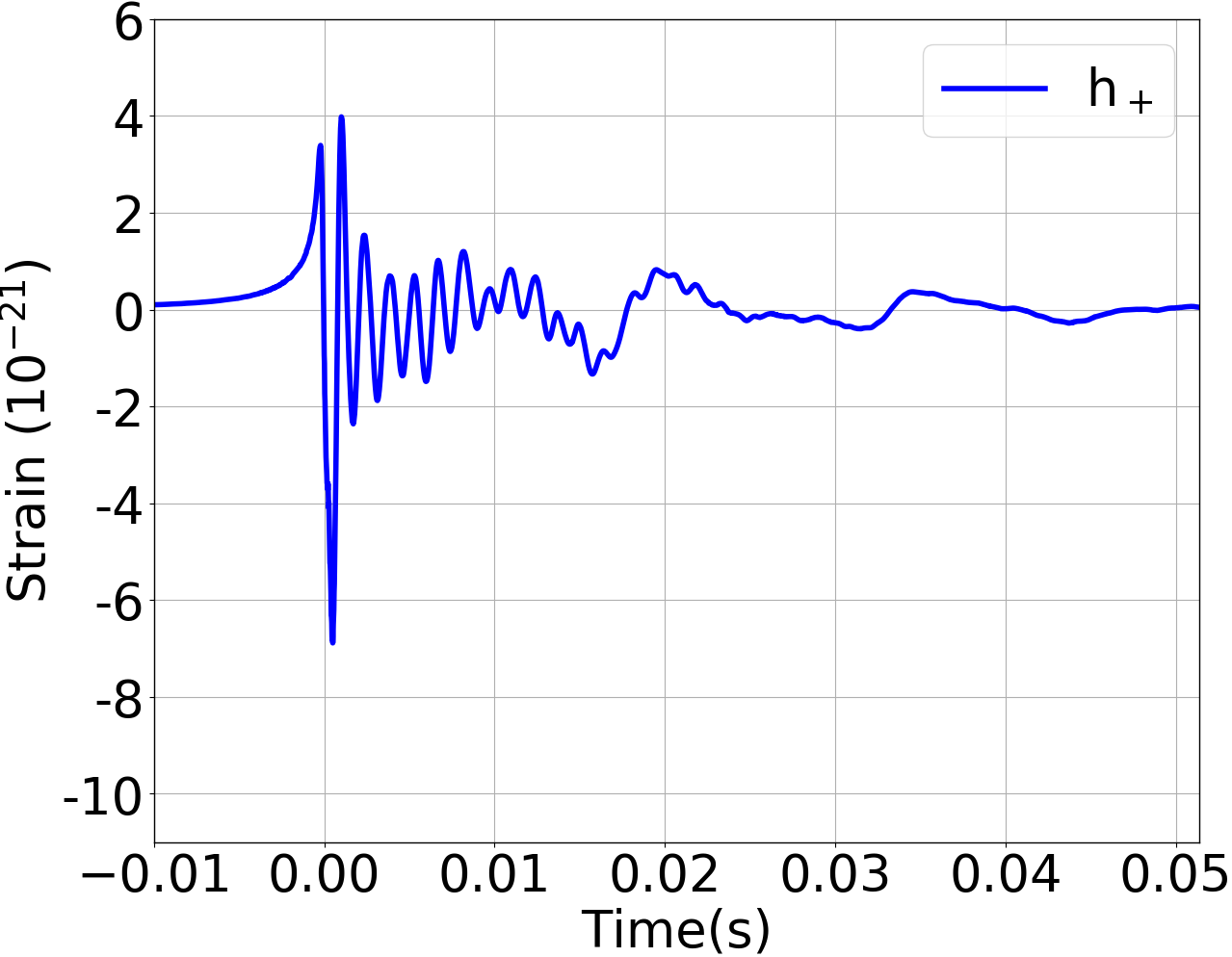}%
        }
    \end{center}
    \caption{Examples of simulated waveforms for which the explosion is modelled by
the magnetorotational mechanism. The left panel shows a waveform
from~\cite{abdikamalov2014measuring}. The progenitor is $12\text{M}_\odot$ with
differential rotation parameter $A=10^4~\text{km}$. The initial angular velocity
at the core $\Omega_c$ is $5.0~\text{rads/s}$. The middle panel shows a waveform
from~\cite{dimmelmeier2008gravitational}. The simulation corresponds to a
progenitor of $15\text{M}_\odot$ with $A=10^5$ km and $\Omega_c =
4.56~\text{rads/s}$.  The right panel shows a waveform
from~\cite{richers2017equation} where the progenitor is $12\text{M}_\odot$, $A$
and $\Omega_c$ are $10^4$ km and $3.0~\text{rad/s}$ respectively. In all panels,
only the $h_{+}$ polarisations are shown because the simulations are
axis-symmetric or 2D, and therefore described by only one polarisation.  The
sources are assumed to be at a distance of $10$ kpc. The x-axes show the time after 
core bounce.\label{fig:magwaveforms}} 
\end{figure*}

\begin{figure*}
     \begin{center}
        \subfigure[]{
            \label{fig:neusample1}
            \includegraphics[width=0.315\textwidth]{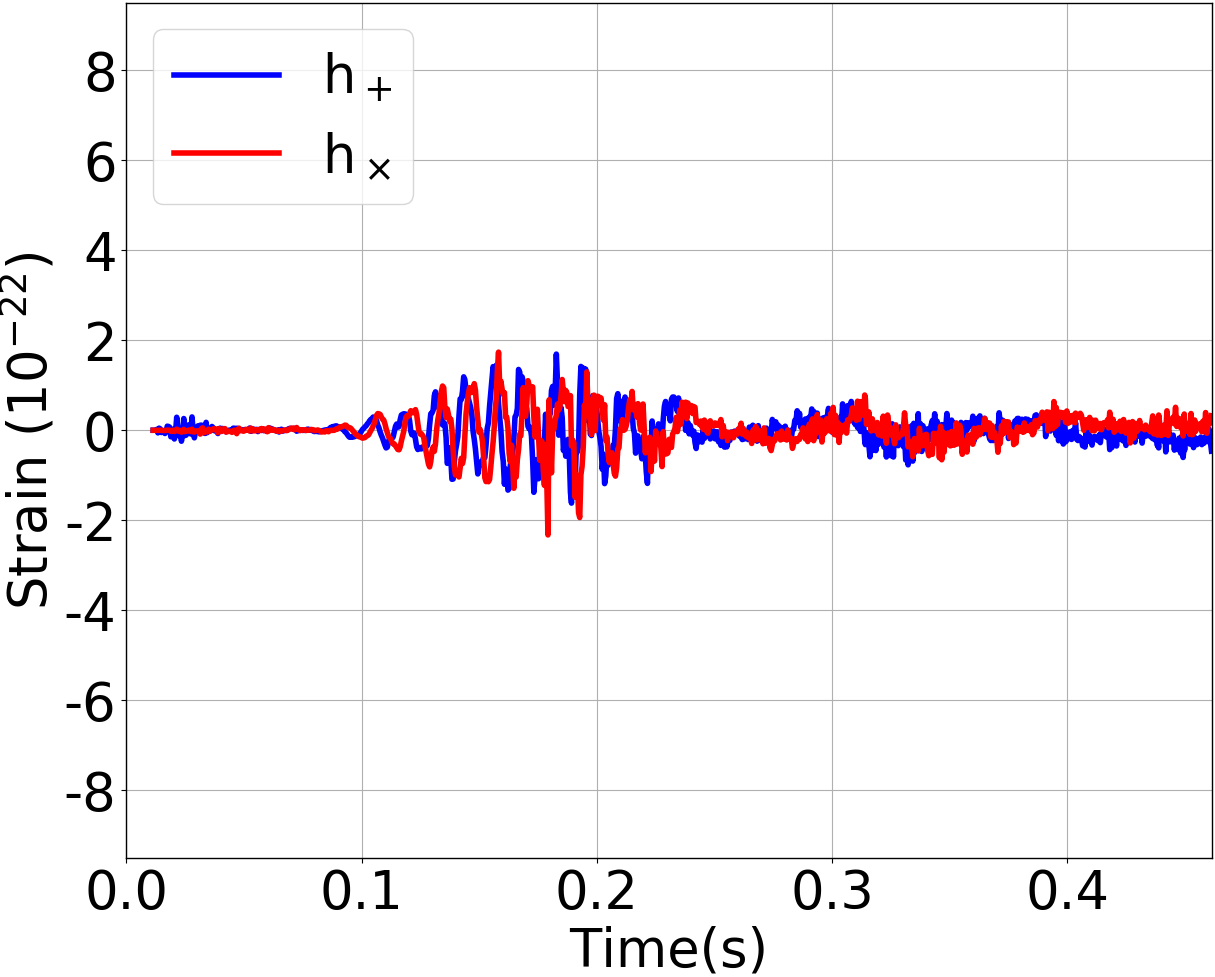}%
        }\quad
        \subfigure[]{
            \label{fig:neusample2}
            \includegraphics[width=0.315\textwidth]{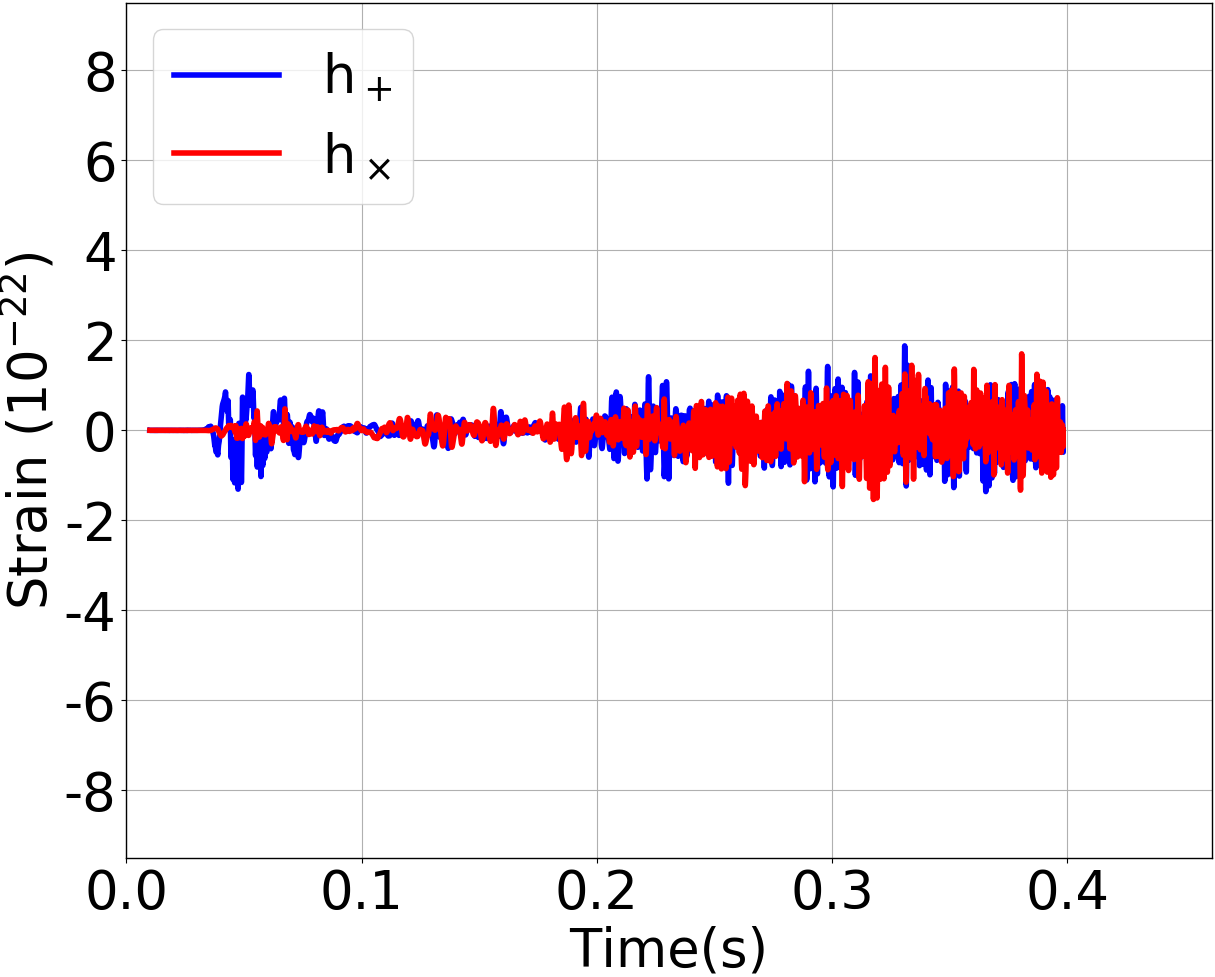}%
        }\quad
        \subfigure[]{
            \label{fig:neusample3}
            \includegraphics[width=0.315\textwidth]{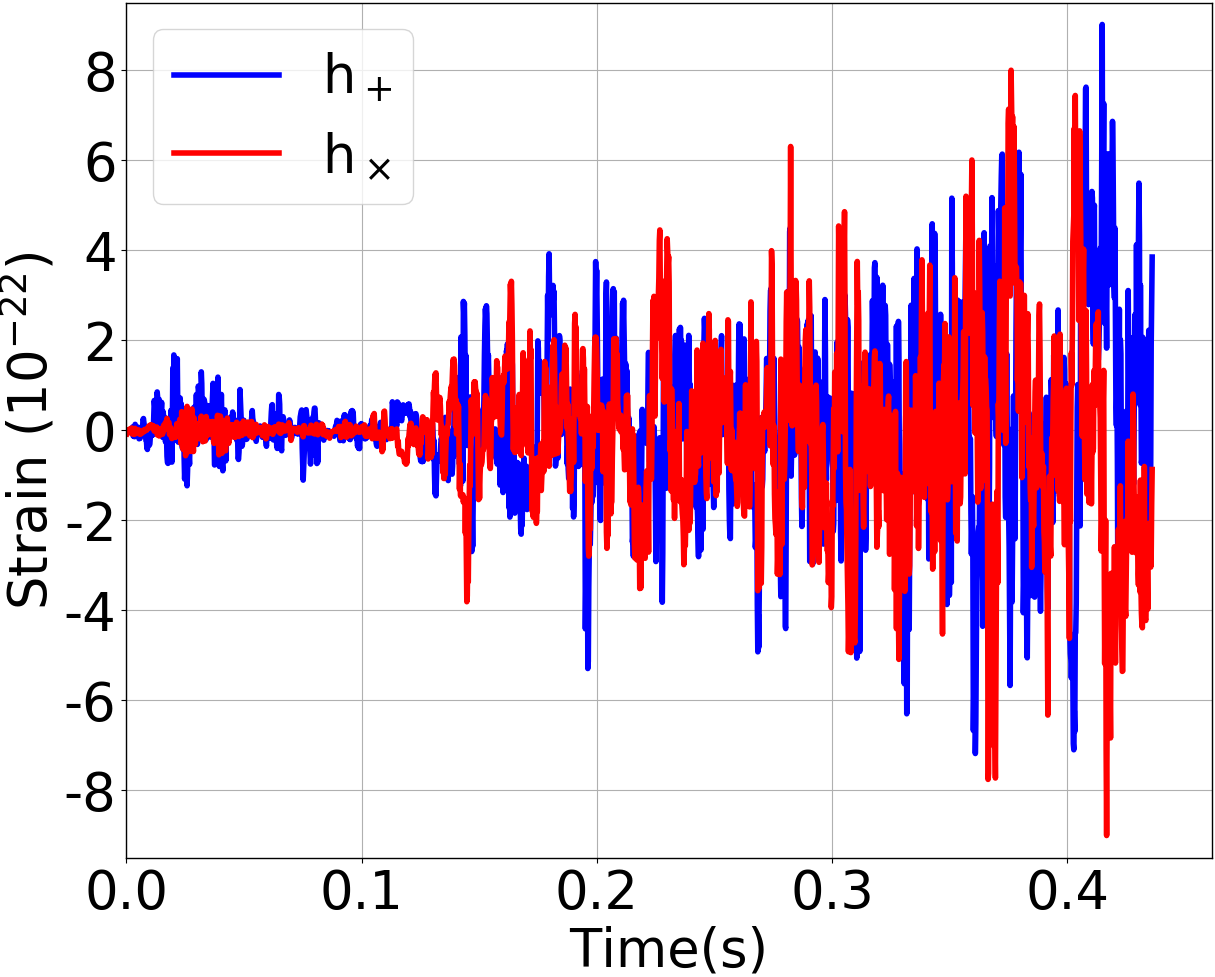}%
        }
    \end{center}
    \caption{Examples of simulated waveforms for which the explosion mechanism is
modelled by the neutrino-driven mechanism. From the left to the right, the
panels show waveforms from~\cite{10.1093mnrasstz990, 
radice2019characterizing, yakunin2017gravitational}, with progenitors of masses equal to
$15\text{M}_\odot$, $15\text{M}_\odot$ and $60\text{M}_\odot$ respectively.
The sources are assumed to be at $10$ kpc from earth. The x-axes show the time after core bounce.
\label{fig:neuwaveforms}} 
\end{figure*}

%
%
For the neutrino-driven mechanism, we employ the waveforms
from~\cite{10.1093mnrasstz990, kuroda2017correlated, muller2012parametrized,
powell2019gravitational, radice2019characterizing, yakunin2015gravitational,
yakunin2017gravitational, ott2009gravitational, murphy2009model,
ott2013general}\footnote{\href{https://wwwmpa.mpa-garching.mpg.de/ccsnarchive/data/Andresen2019/}
{https://wwwmpa.mpa-garching.mpg.de/ccsnarchive/
data/Andresen2019/ }}$^{,}$\footnote{\href{ https://www.astro.princeton.edu/~burrows/gw.3d/  }
{ https://www.astro.princeton.edu/~burrows/gw.3d/  }}. The simulations in~\cite{10.1093mnrasstz990,
kuroda2017correlated, muller2012parametrized, powell2019gravitational,
radice2019characterizing, yakunin2017gravitational, ott2013general} were from 3D modelling of
the supernovae while the simulations in~\cite{yakunin2015gravitational,
ott2009gravitational, murphy2009model} were 2D. These
simulations cover a wide range of progenitor masses from $9\text{M}_\odot$ to
$60\text{M}_\odot$ and the specific progenitor masses of the simulations are
shown in Table~\ref{table:waveforms}. Examples of the waveforms for
both mechanisms are shown in Figs.~\ref{fig:magwaveforms} and
\ref{fig:neuwaveforms}.

%
%
\begin{table}[]
\centering
\begin{threeparttable}
\caption{Waveforms}
\label{table:waveforms}
\begin{tabular}{rccc}
\toprule
    & Mechanism         & Mass ($\text{M}_\odot$)                         & No.  \\
\hline                         
Abdikamalov~\cite{abdikamalov2014measuring} & M   & 12.0                          & 92   \\
Dimmelmeier~\cite{dimmelmeier2008gravitational} & M   & 11.2,15.0,20.0,40.0        & 136  \\
Richers~\cite{richers2017equation}     & M   & 12.0                          & 1824 \\
Andresen~\cite{10.1093mnrasstz990}    & N   & 15.0                          & 6    \\
Kuroda~\cite{kuroda2017correlated}      & N   & 11.2,15.0                    & 2    \\
Muller~\cite{muller2012parametrized}      & N   & 15.0,20.0                    & 6    \\
Murphy~\cite{murphy2009model}      & N   & 12.0,15.0,20.0,40.0        & 16   \\
Ott$_1$~\cite{ott2009gravitational}       & N   & 15.0                          & 2    \\
Ott$_2$~\cite{ott2013general}       & N   & 27.0                          & 8    \\
Powell~\cite{powell2019gravitational}      & N   & $3.5^{*}$,18.0                      & 2    \\
Radice~\cite{radice2019characterizing}      & N   & 9,10,11,12,13,19,25,60 & 8    \\
Yakunin$_1$~\cite{yakunin2015gravitational}   & N   & 12,15,20,25                & 4    \\
Yakunin$_2$~\cite{yakunin2017gravitational}   & N   & 15                            & 1   \\
\hline
\hline
\end{tabular}
\begin{tablenotes}
\setlength\labelsep{0pt}
\normalfont{
\item The mass ranges and mechanisms of the progenitors of the simulated
waveforms used in this work.  The first column refers to the studies. Mechanism
indicates the explosion mechanism for the waveforms, with M for the
magnetorotational mechanism and N for the neutrino-driven mechanism.  Mass
refers to the mass, in units of solar mass, of the progenitors in the simulations.
No. means the number of waveforms available from the study.
All masses are the masses of the stars at
zeros age unless indicated otherwise.\\
$^*$ Mass of a star in a binary system with an initial helium mass of $3.5\text{M}_\odot$} 
\end{tablenotes}
\end{threeparttable} 
\end{table}


%
%
%
%
%
%
Introducing waveforms from a variety of studies introduces a distribution of
the waveforms that covers a larger parameter space, making the data samples
harder for the \ac{CNN} to learn. As a result, the performance of the trained
\ac{CNN} on the testing samples may appear to be worse than that if the data
were generated using only a subset of the waveforms. However, a larger
parameter space provides the advantage of forcing the \ac{CNN} to learn the
features common among the waveforms rather than simply a subset of the
waveforms.  The trained network can thus perform better when presented with
waveforms with unexpected features, which is more likely in reality.

%
%
Since the waveforms are generated at various distances, sampling rates and
durations, it is necessary to normalise the waveforms before they can be used
for the generation of the time-series. To do this, we first scale the
amplitudes of the waveforms by moving the sources to $10$ kpc from earth. We
then ensure that the sampling rate are identical for all the waveforms by down
sampling them to a pre-selected sampling rate of $4096$Hz. The longest duration
$\tau$ among the waveforms is then identified and each of the remaining
waveforms is padded with zeros to this duration. To introduce as few artefacts
as possible, a high pass filter with a low cut-off frequency equal to $11$Hz
and a tukey window ($\alpha = 0.08$) are applied prior to the zero padding. 

%
%
To balance the difference in the number of waveforms between the two mechanisms
in the training, validation and testing data set, $36$ exact copies of each neutrino-driven waveform are
made. This will make the ratio of the waveforms modelled by the two mechanisms close to 1, 
while does not change the ratio of the waveforms from different studies modelled by the neutrino-driven mechanism. 
This means each waveform in the neutrino-driven mechanism will have equal representations in the data set,
and waveforms from different mechanisms will have equal representation.
After this procedure, the simulated waveforms are then $\bold{S}(t) = \{\bold{s}_{1}(t), \bold{s}_{2}(t),
..., \bold{s}_{m}(t)\}$, where $m$ is the number of the waveforms and
$\bold{s}_{q}$ is the $q^{\text{th}}$ waveform, defined by, 
\begin{equation}\label{eq:ht}
\bold{s}_q(t) = \left(\begin{array}{c} h_q^+(t) \\ h_q^\times(t)
\end{array}\right), 
\end{equation} 
where $h_q^+(t)$ and $h_q^\times(t)$ are the two polarisations of the waveform
and $t$ is the time. In this work, we assume a 
fixed GPS time when generating the training/validation/testing data. 
At this GPS time, the value of $t$ is zero.
As mentioned, some of the waveforms used in the work are
generated with the simulations being axisymmetric, in which case the waveforms
are entirely described by one polarisation $h_q^+(t)$. The corresponding
$h_q^\times(t)$ for these waveforms are defined as a vector of zeros. 
%
%
In this work, we perform simulations for distances equal to $10$, $20$, $30$,
$40$, $50$, $60$, $80$, $100$, $150$ and $200$ kpc. If for a training
procedure, the default distance $d_\text{L}$ of a waveform is not $10$ kpc, the amplitude are rescaled 
by the distance since the amplitude is inversely proportional to the distance.

%
%
The next step is to generate simulated time-series for the \ac{GW} detectors in
a network using $\bold{S}(t)$.
Since the purpose of building a \ac{CNN} is to categorise an input data sample
into three exclusive classes, in total three types of time-series are
generated.  For the time series containing either a magnetorotational signal or
a neutrino-driven signal, we start by selecting a waveform $\bold{s}_q$ from
$\bold{S}(t)$. A random location of (right ascension, declination) $=(\alpha,
\delta)$ in the sky is selected from a uniform distribution on $\alpha$ and a
uniform distribution on $\sin\delta$. 
%
%
The relative delays in the arrival times of the signals at each
detector are a function of the sky location and are also computed and applied
to the selected waveform. The delay in arrival time between a detector and the
centre of the earth is given by,
\begin{equation}\label{eq:tdetector}
\Delta t = \frac{\bold{n}(\alpha,\delta) \cdot \bold{r}}{c},
\end{equation}
where $\bold{n}(\alpha,\delta)$ is the propagation direction of the \ac{GW},
$c$ is the speed of light, and $\bold{r}$ the location vector of the detector
relative to the centre of the Earth. The resulting $j^{\text{th}}$ signal $\bold{h}_j(\alpha,
\delta, t)$ received by the detector network is then described by,
\begin{equation}\label{eq:ht2}
 \bold{h}_j(\alpha, \delta, t) = \bold{F}(\alpha, \delta, t) \times
\bold{s}_q(t+\bold{\Delta t}),
\end{equation}
where $\bold{F}(\alpha, \delta, t)$ is a matrix of which the elements are the antenna patterns of each detector in the network,
\begin{equation}\label{eq:ap}
\bold{F}(\alpha, \delta, t)= \left(\begin{array}{cc}
f_1^+(\alpha, \delta, t)~~f_1^\times(\alpha, \delta, t) \\
\vdots~~~~~~~~~~~~\vdots \\
f_k^+(\alpha, \delta, t)~~f_k^\times(\alpha, \delta, t) \\
\end{array}\right),
\end{equation}
$f^+(\alpha, \delta, t)$ and $f^{\times}(\alpha, \delta, t)$
are the antenna pattern functions for the two polarizations and $k$ is the
number of detectors as defined above. 
%
%
Next, we generate independent Gaussian noise $\bold{N}_j(T) =
\{\bold{n}_{1j}(T), \bold{n}_{2j}(T), ..., \bold{n}_{kj}(T)\}'$ for each
detector in the network using their respective power spectral
densities. In the expression, $T$ is the time for the generated noise and should not be confused with the lower case $t$, 
which is the time for the waveforms.
To make the simulated data as realistic as possible, 
we allow the start time of the signal in a data sample to vary.
To do this,  the duration of the generated noise $T_{\text{max}}$ is $1.6$ times longer than
that of $\bold{h}_j(\alpha, \delta, t)$. A random number $p$ drawn from 
a uniform distribution on the range of $[0, P)$ 
will then be generated, where
$P = T_{\text{max}}\times 18\% $  and $p$ 
determines where in
the generated noise the signal will be placed, as given by,
\begin{equation}
\bold{d}_j(\alpha, \delta, T, p)=
\\
\begin{cases}
\bold{N}_j(T)~~~~~~~~~~~~~~~~~t < T_{\text{max}} p; \\
\bold{N}_j(T) + \bold{h}_j(\alpha, \delta, t)~ T_{\text{max}} p \leq T \leq \tau + T_{\text{max}} p; \\
\bold{N}_j(T)~~~~~~~~~~~~~~~~~t > \tau + T_{\text{max}} p, \\
\end{cases}
\end{equation}
where $\tau$ is defined in section \ref{sec:spwf}. 
This will avoid the possibility that the \ac{CNN}
learns the human artefact instead of common features of the waveforms by having
the signals always starting at the same place.
For each time-series of the background noise class, independently simulated Gaussian
background noise  of the
same duration as that of the other classes are generated using the same power spectrum density 
for each detector in the network. This means a data sample for this class is defined as,
\begin{equation}\label{eq:noise}
 \bold{d}_j(T) = \bold{N}_j(T).
\end{equation}

\begin{figure}
\includegraphics[width=0.475\textwidth]{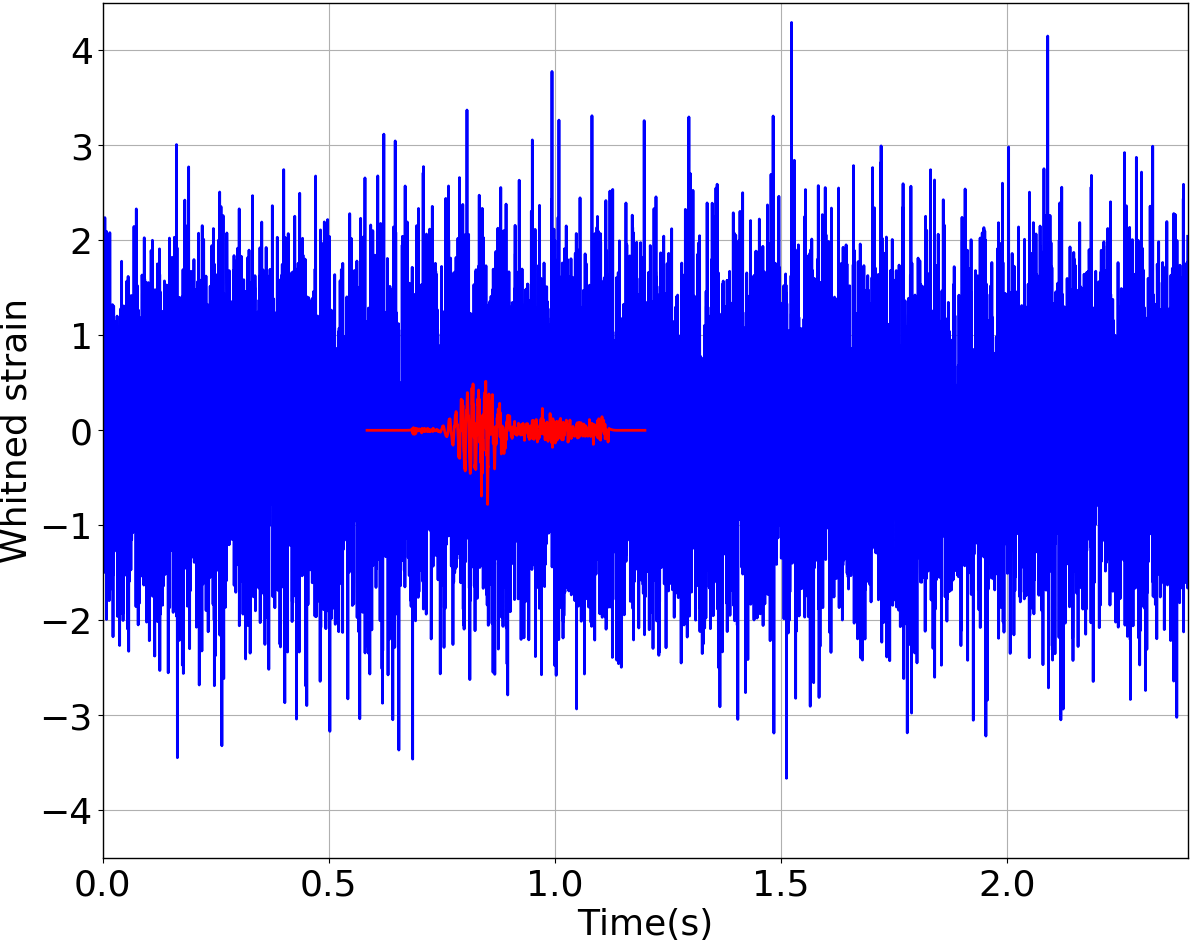}
\caption{A representative example of the simulated time-series used to
train/validate/test the \ac{CNN}. The blue shows a whitened time-series with a
signal added to Gaussian noise generated using the power spectral density of \ac{aLIGO}. 
The red curve shows the same whitened signal free of
noise. The original waveform is shown in Fig. \ref{fig:neusample1}.
The source is at a distance of $10$ kpc from earth.
\label{fig:sample}} 
\end{figure}

%
%
As a general rule, the performance of a \ac{CNN} can always be enhanced by
increasing the volume of training data. Therefore we augment our data set by
iterating over $\bold{S}(t)$ until for each waveform in $\bold{S}(t)$, we have
$31$ data samples $\bold{d}$ generated using the procedure described above.
This step is essential for the \ac{CNN} to learn the features of the waveforms
and to identify them effectively under different noise scenarios and starting
times.  After that, we generate independent noise realisations for the
background noise class. The entire data set $\bold{D}$ has
roughly $l = 1.8 \times 10^{5}$ data samples, where each of the 3 classes
contains approximately $6\times10^{4}$ data samples.
%
%
%
%
In Fig.~\ref{fig:sample}, we show a representative example of a pre-processed
input time-series for a single detector. A data sample consists of such a
time-series containing a signal from the same source or simply background noise
from all the detectors in the network. The data samples are then split into 3
groups, with $1\times10^{4}$ samples being randomly selected for validation,
$10\%$ of the remainder for testing and $90\%$ for training. When the training
is finished for a dataset generated at a given distance, the above described
procedure will be repeated for another distance with a different \ac{CNN} of 
the same architecture until the training for all
distances have been carried out for a \ac{GW} detector network. The entire
procedure is then repeated for another \ac{GW} detector network. 
As mentioned in the introduction, the \ac{CNN} will be trained for the two networks of
\ac{GW} detectors presented in Table~\ref{table:network}. For the remaining of
the paper, we will use their acronyms to refer to the networks, i.e., HLVK and H+L+VK.

%
%

\begin{table}[]
\centering
\begin{threeparttable}
\caption{Detector networks}
\label{table:network}
\begin{tabular}{lll}
\toprule
Network            & \multicolumn{1}{c}{Detector}                & Acronym                 \\
\hline
\multirow{4}{*}{1} & \ac{aLIGO} Hanford~\cite{aasi2015advanced}    & \multirow{4}{*}{HLVK}   \\
                   & \ac{aLIGO} Livingston &                         \\
                   & \ac{AdVirgo}~\cite{acernese2014advanced}           &                         \\
                   & KAGRA~\cite{aso2013interferometer}                    &                         \\
                   \hline
\multirow{4}{*}{2} & LIGO A+ Hanford~\cite{miller2015prospects, LIGOW}          & \multirow{4}{*}{H+L+VK} \\
                   & LIGO A+ Livingston       &                         \\
                   & \ac{AdVirgo}           &                         \\
                   & KAGRA                    &                         \\
\hline
\hline
\end{tabular}
\begin{tablenotes}
\setlength\labelsep{0pt}
\normalfont{
\item The networks of detectors used in this work.}
\end{tablenotes}
\end{threeparttable}
\end{table}

\section{Result and discussion}\label{sec:result}
%
%
After the \ac{CNN} is trained, we can estimate its performance using the
testing samples and the results of which are presented in this section. 
By applying the trained network to the testing samples, 
the \ac{CNN} will output values of statistics to each of the testing samples. 
Since we have knowledge of the true class associated with the testing samples,
it is possible to construct the \ac{ROC} curves.
The \ac{ROC} curve is one of the most commonly used and convenient ways to 
determine the classification performance of a \ac{CNN} or equivalent signal detection 
algorithm. A \ac{ROC} shows the performance of a classifying model by
defining the \ac{TAP} as a function of the \ac{FAP}. Since \ac{ROC} curves are
usually plotted for models distinguishing between two classes, for a
multi-class classification problem, a \ac{ROC} for a class should be viewed as
the class versus the others. This means in this context, \ac{FAP} means the
fraction of samples from other classes misidentified as a sample from the class
the \ac{ROC} is associated with. The \ac{TAP} is identical to that of a
two-class classification problem and indicates the fraction of samples
correctly identified.  For a given \ac{FAP}, a model with a higher \ac{TAP} is
considered more capable than a model with a lower \ac{TAP}.

\begin{figure*}
     \begin{center}
        \subfigure[]{
            \label{fig:ROClog1}
            \includegraphics[width=0.475\textwidth]{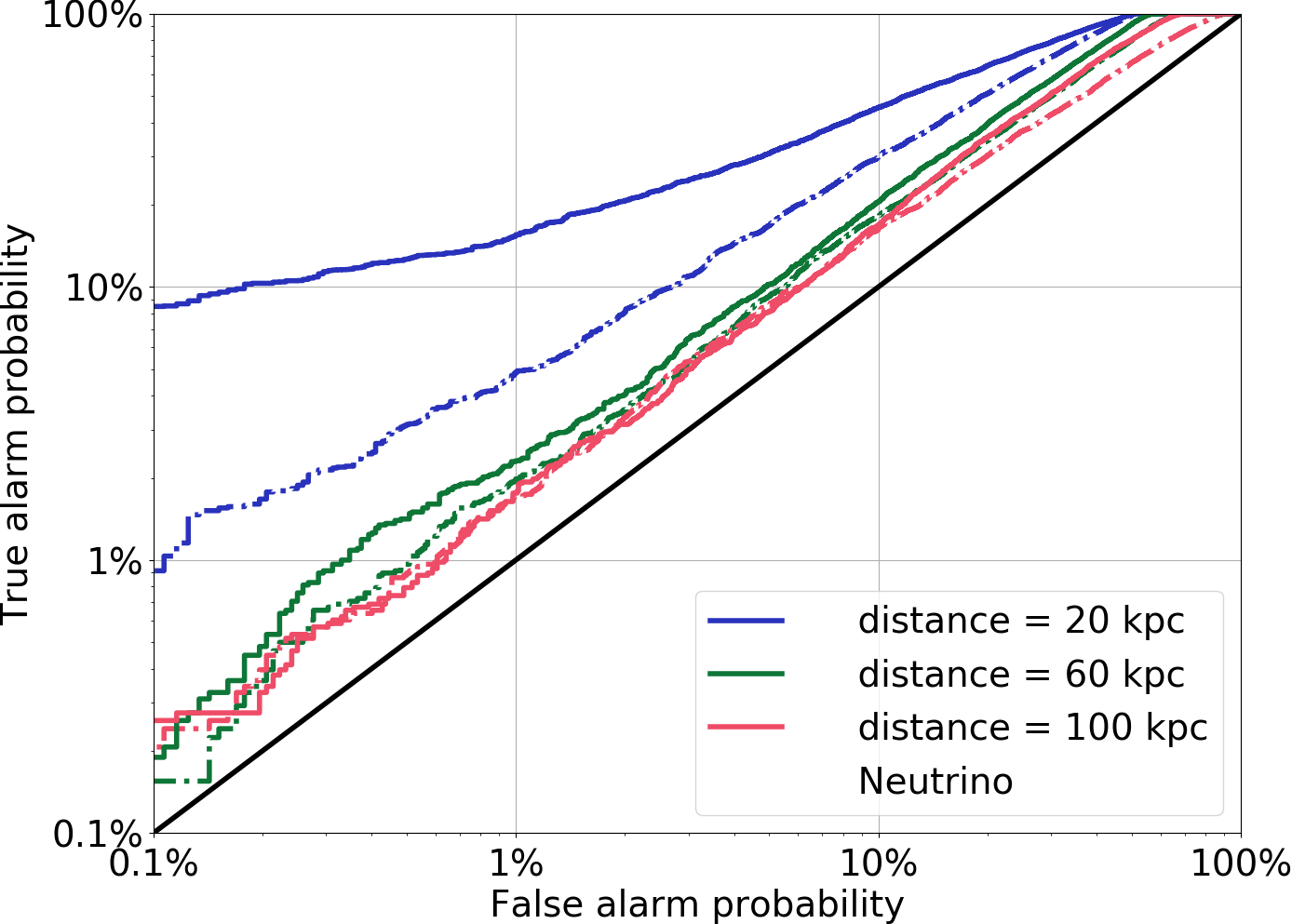}%
        }\quad
        \subfigure[]{
            \label{fig:ROClog2}
            \includegraphics[width=0.475\textwidth]{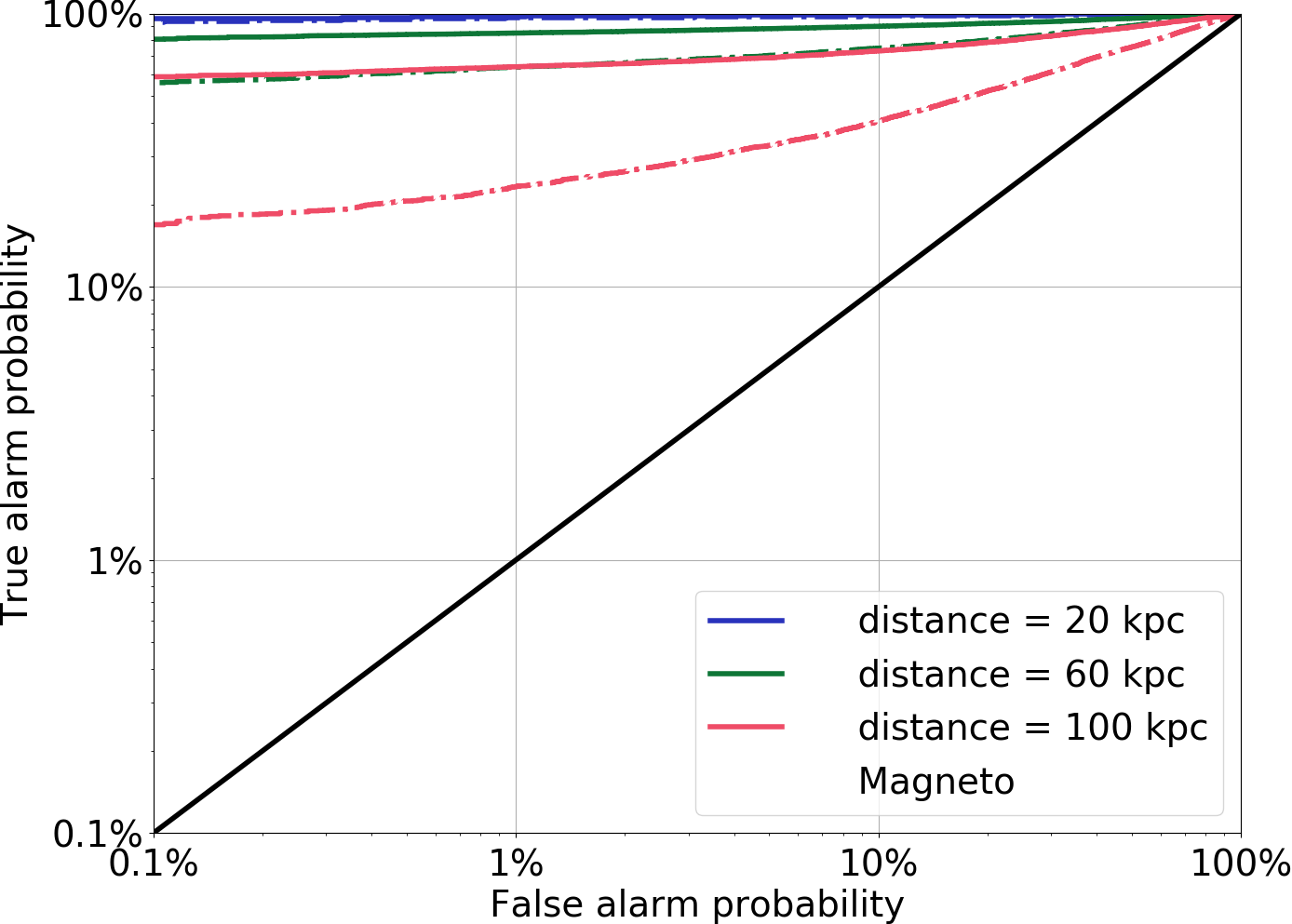}%
        }
    \end{center}
    \caption{\ac{ROC} curves showing the classification performance of the
\ac{CNN} for \ac{CCSN} signals of different explosion mechanisms at three
distances, $20$, $60$, and $100$ kpc. The left panel shows the results for the
neutrino-driven mechanism, while the panel on the right shows those for the
magnetorotational mechanism. In both panels, the solid lines are for the
network H+L+VK, while the dashed lines are for HLVK. The black diagonal lines in both plots indicate 
the worst scenario where a model has zero ability in distinguishing its input.
\label{fig:ROClog}} 
\end{figure*}

%
%
In Fig.~\ref{fig:ROClog}, we show the \acp{ROC} for both of the \ac{CCSN}
mechanisms and \ac{GW} detector networks tested in this work. For simplicity,
we show only the results for three distances, namely, $20$, $60$ and $100$ kpc.
For all the distances tested, the \ac{CNN} achieves a higher \ac{TAP} for any
given \acp{FAP} for magnetorotational than neutrino-driven signals.  That is
not surprising as the intrinsic amplitudes for magnetorotational signals are
higher than those of neutrino-driven signals.

\begin{figure*}
     \begin{center}
        \subfigure[]{
            \label{fig:eff1}
            \includegraphics[width=0.475\textwidth]{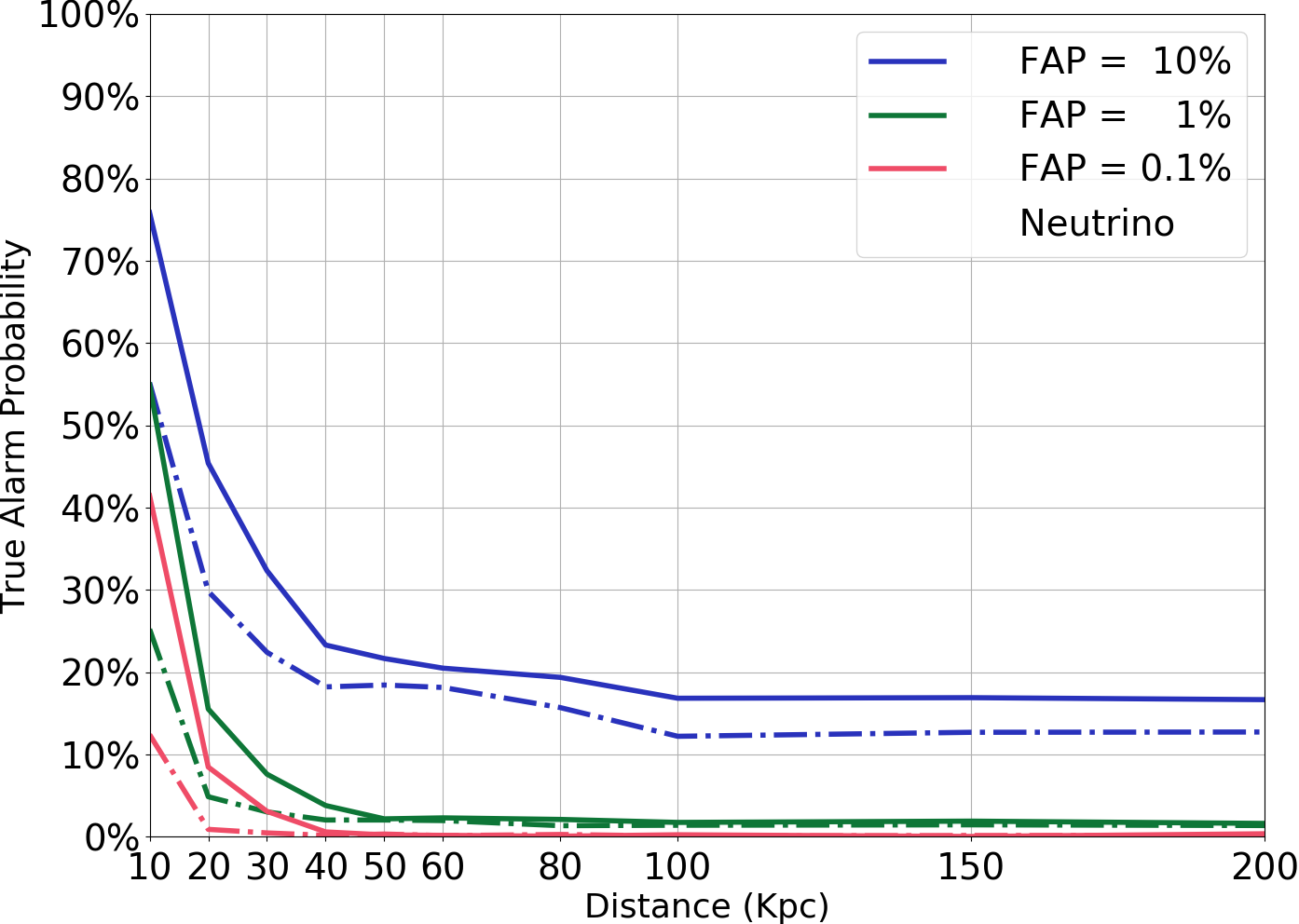}
        }\quad
        \subfigure[]{
            \label{fig:eff2}
            \includegraphics[width=0.475\textwidth]{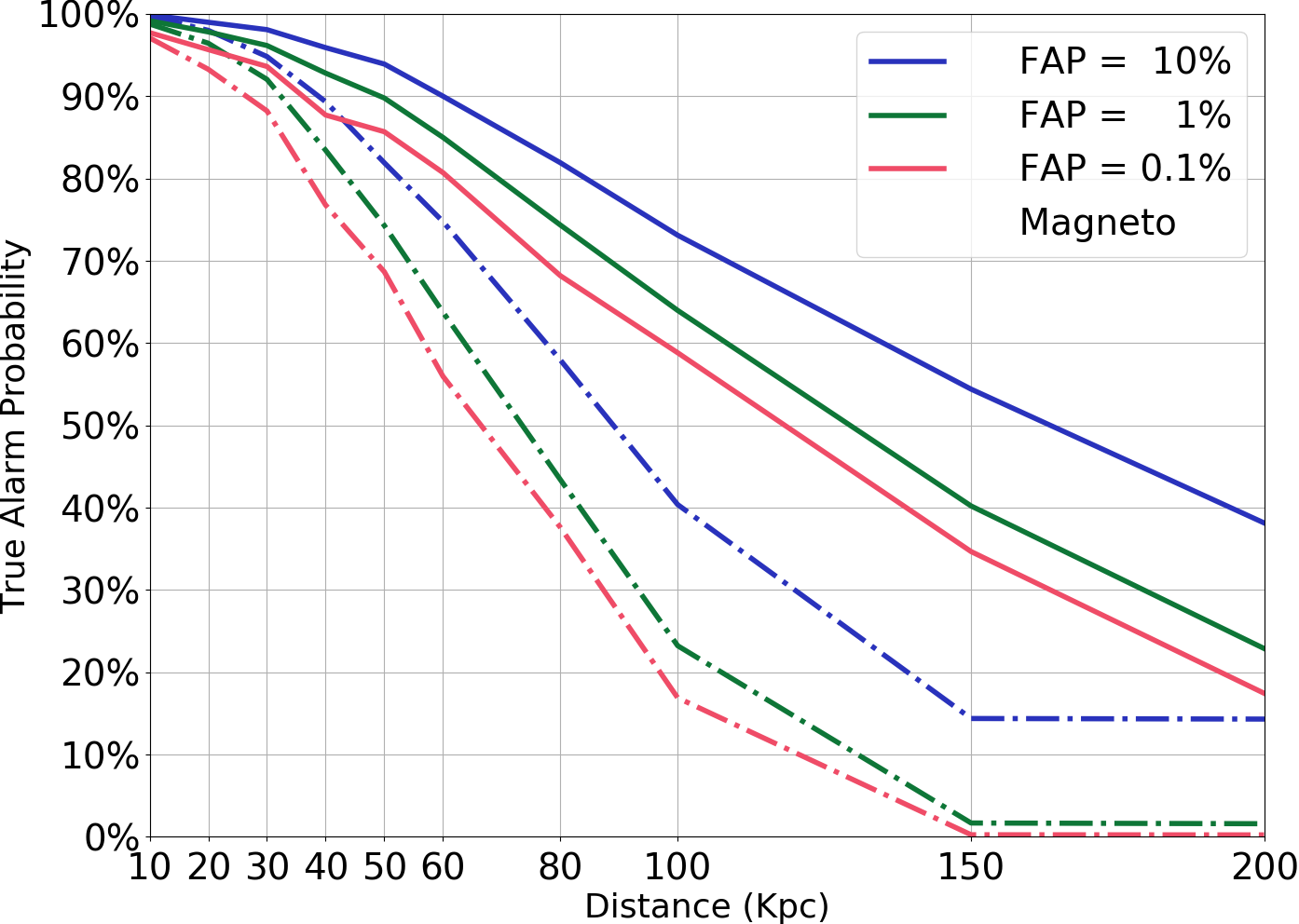}
        }
    \end{center}
    \caption{Efficiency curves showing the classification ability of the
\ac{CNN} as a function of distance for both mechanisms and networks.  The left
panel shows the results for the neutrino-driven mechanism, while the right
shows those for the magnetorotational mechanism. In both panels, the solid
lines show the results for the network H+L+VK, and the dashed lines for HLVK.
Three \acp{FAP} values are chosen, blue for \ac{FAP} $=10\%$, green for \ac{FAP}
$=1\%$, red for \ac{FAP} $=0.1\%$.\label{fig:eff}}
\end{figure*}

%
%
We also show the classification efficiency of the \ac{CNN}  as a function of
distance.  This is done by fixing the \ac{FAP} and plotting the \ac{TAP}.  The
results for three chosen \acp{FAP} values are shown in Fig.~\ref{fig:eff}.  In
this figure, a similar trend is seen that magnetorotational signals are easier
for the \ac{CNN} to identify than neutrino-driven signals at a given distance.
For magnetorotational signals from sources located at $50$ kpc and a \ac{FAP}
of $10\%$, the \ac{CNN} achieves a \ac{TAP} of $94\%$ and $82\%$ for the
networks H+L+VK and HLVK respectively. At more restrict \acp{FAP} such as
$0.1\%$, the \ac{CNN} still achieves \acp{TAP} $86\%$ and $68\%$ for sources at
the same distance for the two networks respectively.  For sources that are
located at a slightly further distance of $60$ kpc, both networks achieve a
\ac{TAP} of close to or larger than $80\%$ at a \ac{FAP} of $10\%$. Such a range
includes distances associated with the Large and Small Magellanic Clouds and
covers the satellite galaxies in between~\cite{karachentsev2004catalog,
belokurov2007cats}.  For sources at $100$ kpc, the \acp{TAP} are close to or
larger than $60\%$ for H+L+VK for all chosen \acp{FAP} values, and is $73\%$ if
the \ac{FAP} is $10\%$. Even for sources at $150$ and $200$ kpc, the \acp{TAP}
are $54\%$ and $38\%$ respectively for the same \ac{FAP}, indicating that with
such a \ac{GW} network, it is possible to detect magnetorotational \ac{CCSN}
signals out to such a distance. On the other hand, it is more difficult for the
\ac{CNN} to detect and classify the neutrino-driven signals, due to their
weaker amplitudes. Nonetheless, for sources at $10$ kpc, the \ac{CNN} achieves
a \ac{TAP} of $76\%$ and $55\%$ for H+L+VK and HLVK respectively if the
\ac{FAP} is $10\%$.  This means that for a \ac{GW} from a Galactic \ac{CCSN}
signal it is possible to detect and classify it with either of these \ac{GW}
detector networks.

%
%

%
%
\begin{figure*}
     \begin{center}
        \subfigure[]{
            \label{fig:ROCfixed1}
            \includegraphics[width=0.475\textwidth]{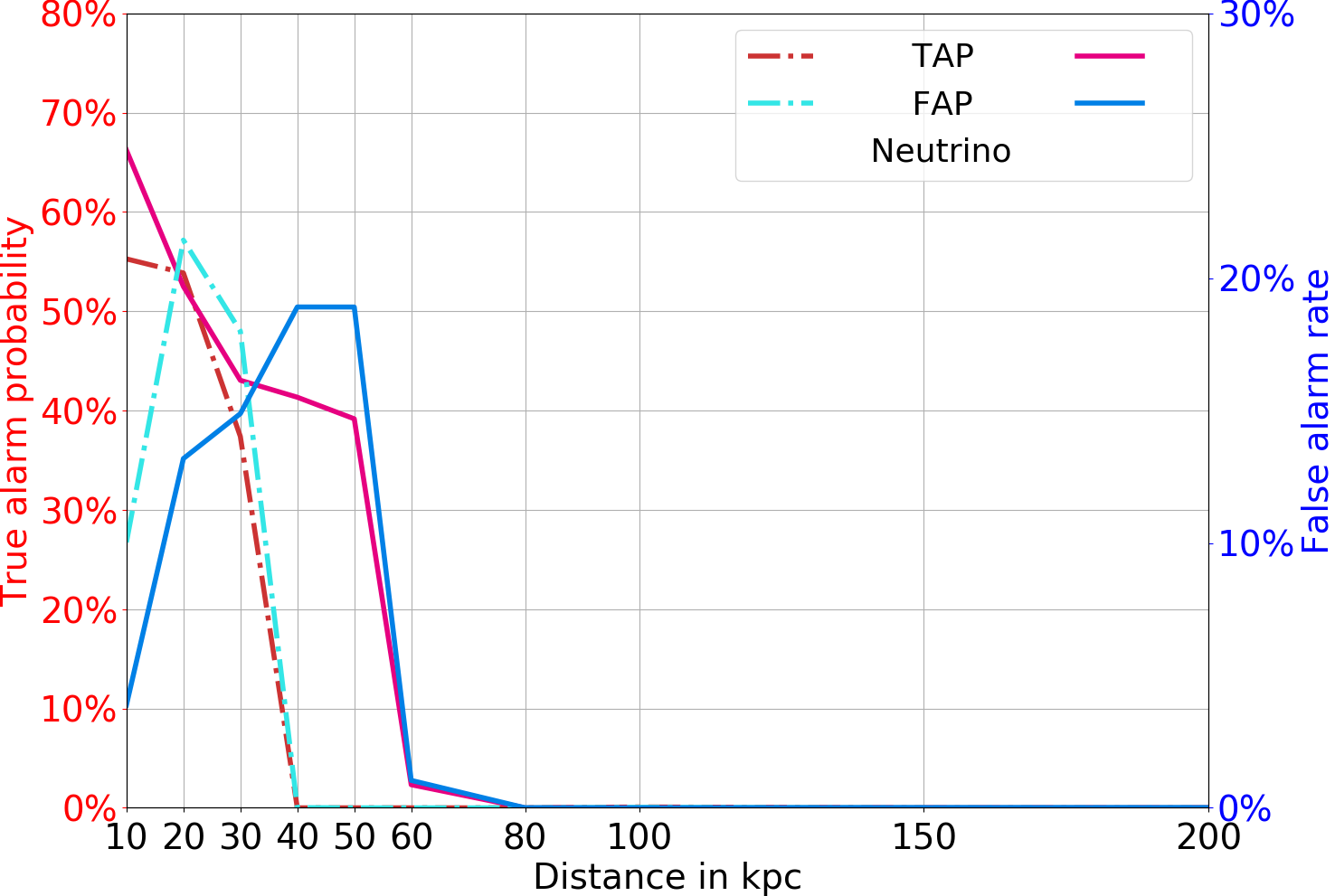}
        }\quad
        \subfigure[]{
            \label{fig:ROCfixed2}
            \includegraphics[width=0.475\textwidth]{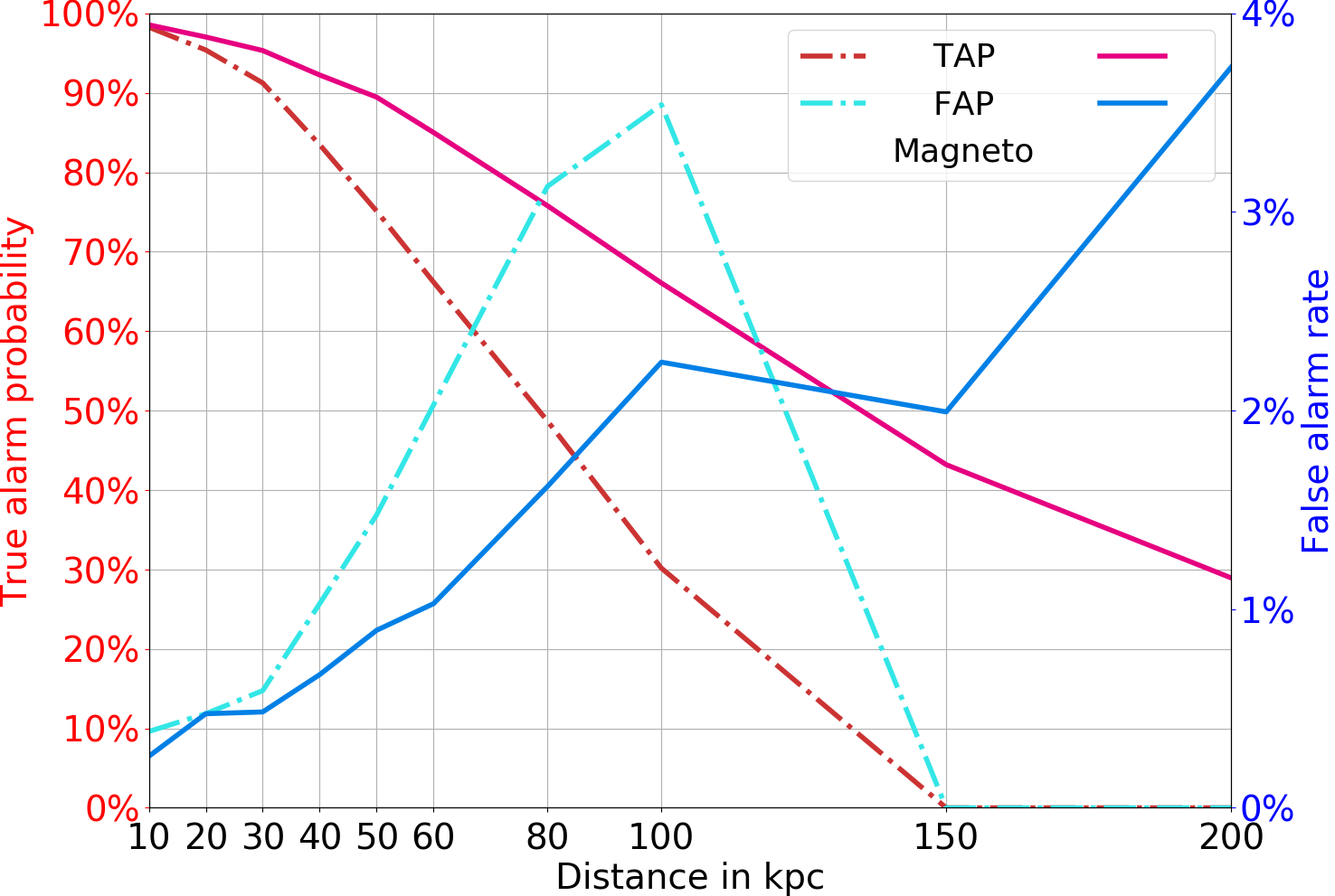}
        }
    \end{center}
    \caption{Efficiency curves showing the ability of the \ac{CNN} in
distinguishing input data with a fixed decision threshold $(50\%)$, 
and their corresponding \acp{FAP} and \ac{TAP}. 
The left panel shows the results for the neutrino-driven mechanism, and
the right shows that for the magnetorotational mechanism. In both panels, the
solid lines show the \acp{TAP} and \acp{FAP} for the network H+L+VK, and the
dashed lines show those for HLVK.\label{fig:ROCfixed}}
\end{figure*}

In practice, since the output of a \ac{CNN} is probabilities indicating how
likely the input belongs to each of the classes, it may be desired to set a
threshold probability on which the decision whether the input belong to a class is made.
For example, an input will be assigned into a certain class if the
corresponding probability is larger than the pre-selected threshold probability. In such a
scenario, the \ac{FAP} and \ac{TAP} would be affected by the choice of the
threshold. We show such a result in Fig.~\ref{fig:ROCfixed}, where we employed
a threshold probability of $50\%$. This value is chosen because for lower values, 
there could be more than a class with predicted probabilities from the \ac{CNN} 
larger than the threshold, while higher values could potentially rule out inputs 
that may otherwise be correctly identified.
For H+L+VK and magnetorotational signals at $10$ kpc, the \ac{TAP} is $99\%$. If the distance
is extended to $80$ kpc, the \ac{TAP} is still close to $80\%$.  For the
largest distances tested in this work, $150$ and $200$ kpc, the \acp{TAP} are
$43\%$ and $29\%$ respectively. 
For HLVK, the \ac{TAP} is $98\%$ and $49\%$ for sources at $10$ and
$80$ kpc respectively.
For all distances, the \ac{CNN} maintains a \ac{FAP} no larger than $4\%$ for
both networks regarding magnetorotational waveforms. For neutrino-driven
signals, we show that H+L+VK has a \acp{TAP} of $66\%$ at $10$ kpc while it
is $55\%$ at $10$ kpc for HLVK.  Both of the networks have \acp{FAP} close to
or less than $20\%$. It should be noted that the results presented in this section are
averaged over all the data samples in the testing samples. The performance on
any individual waveform may vary depending on the morphology and the amplitudes
of the waveform.


\section{Waveforms Excluded from Training}\label{sec:unseen}
\begin{figure*}
     \begin{center}
        \subfigure[]{
            \label{fig:ex_waveform_mag1}
            \includegraphics[width=0.475\textwidth]{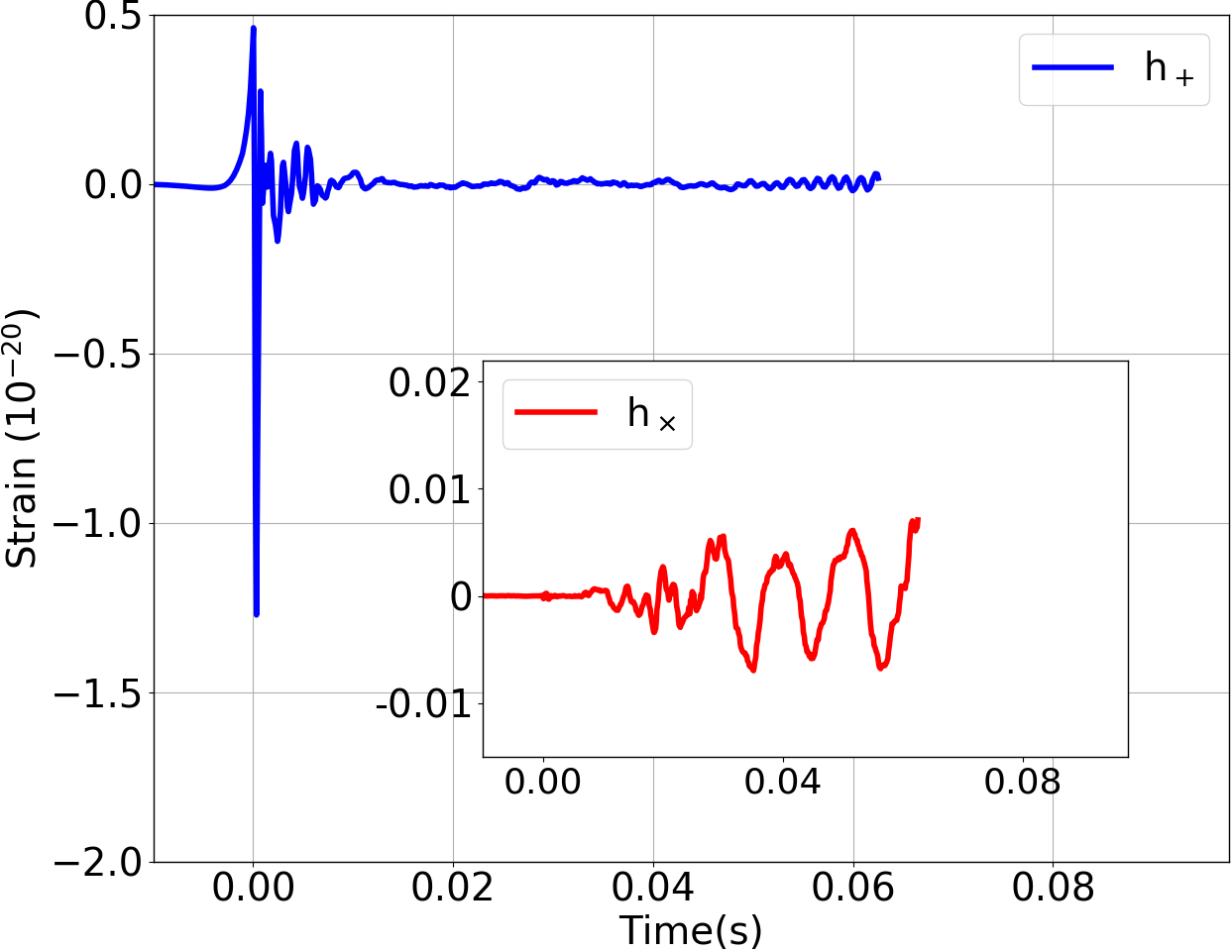}
        }
        \subfigure[]{
            \label{fig:ex_waveform_mag2}
            \includegraphics[width=0.475\textwidth]{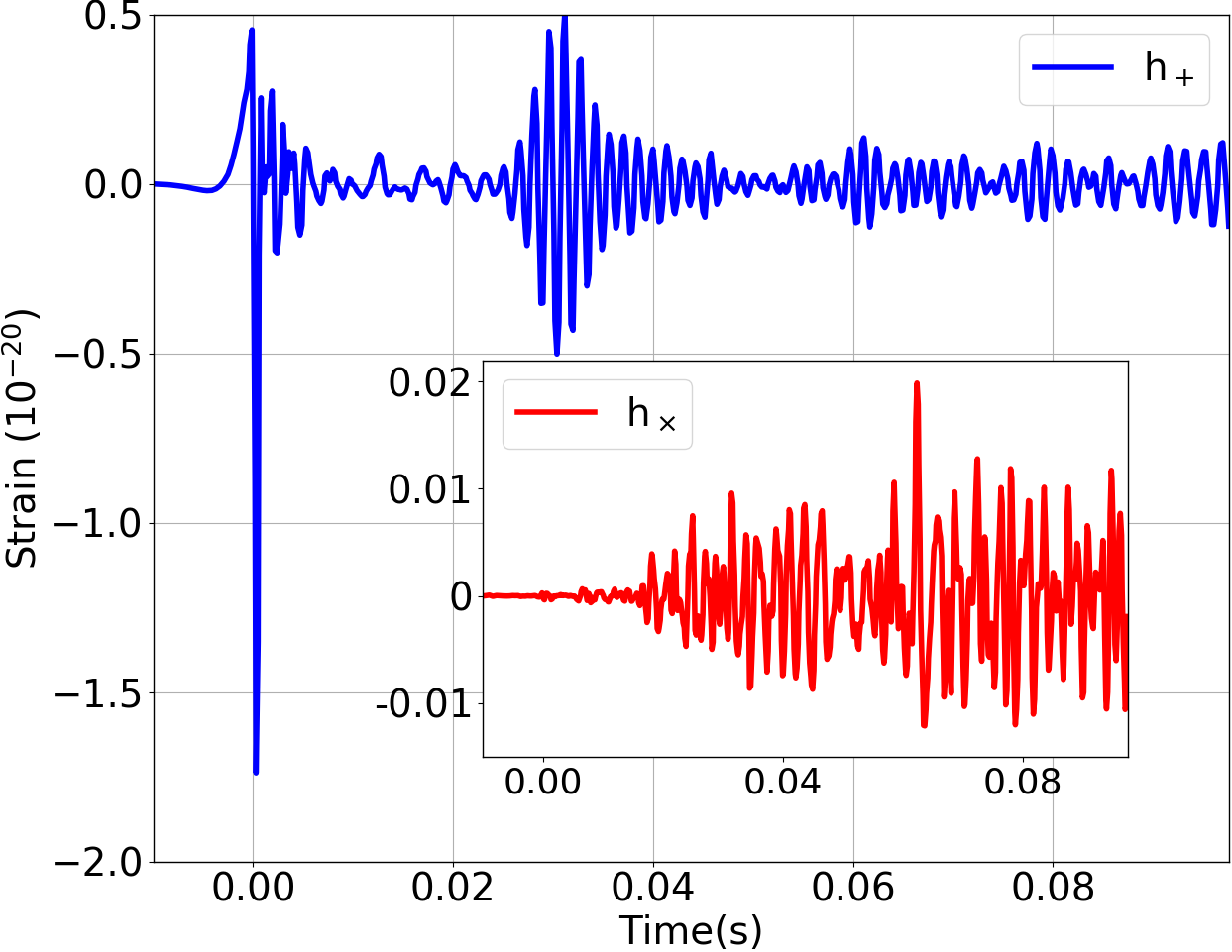}
        }\quad
        \subfigure[]{
            \label{fig:ex_waveform_neu1}
            \includegraphics[width=0.475\textwidth]{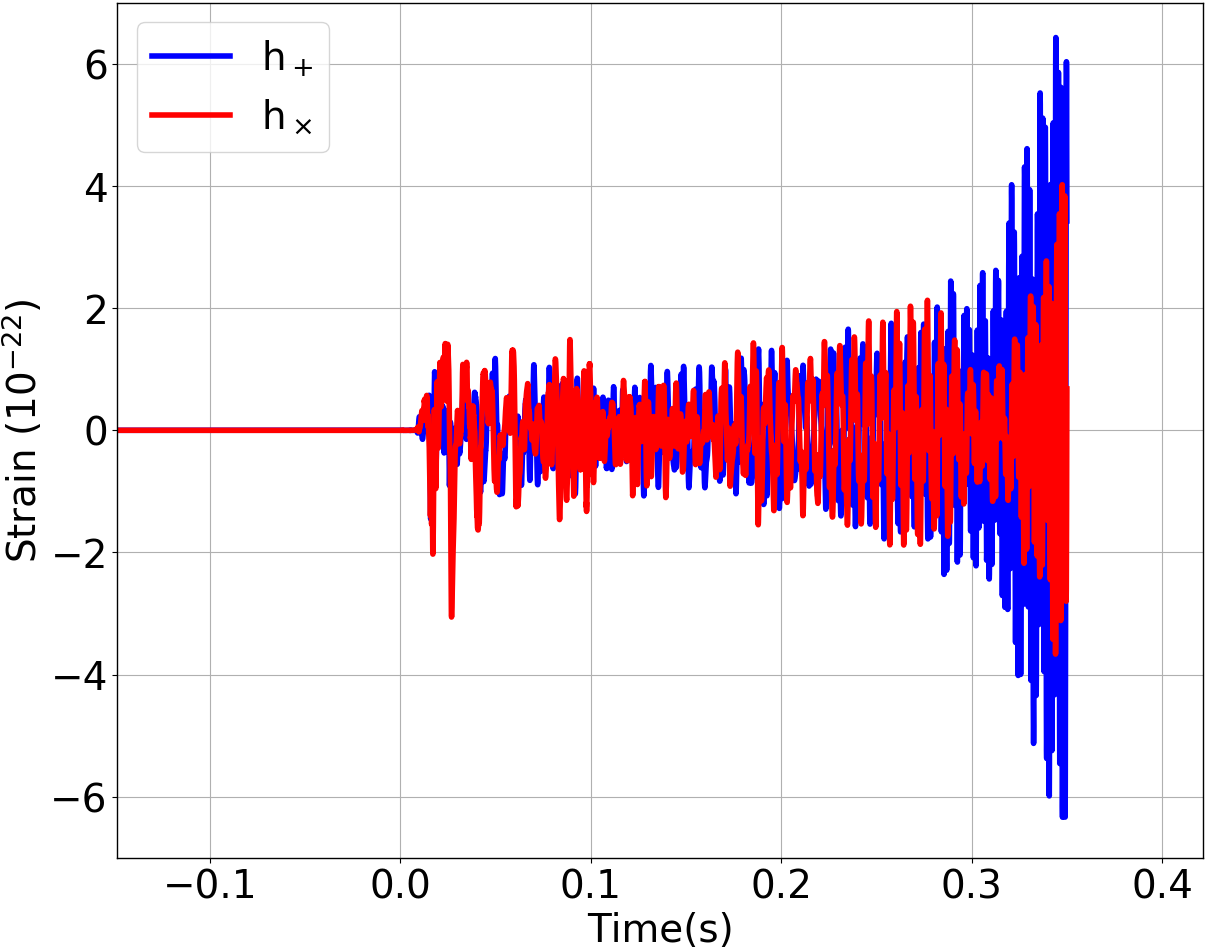}
        }
        \subfigure[]{
            \label{fig:ex_waveform_neu2}
            \includegraphics[width=0.475\textwidth]{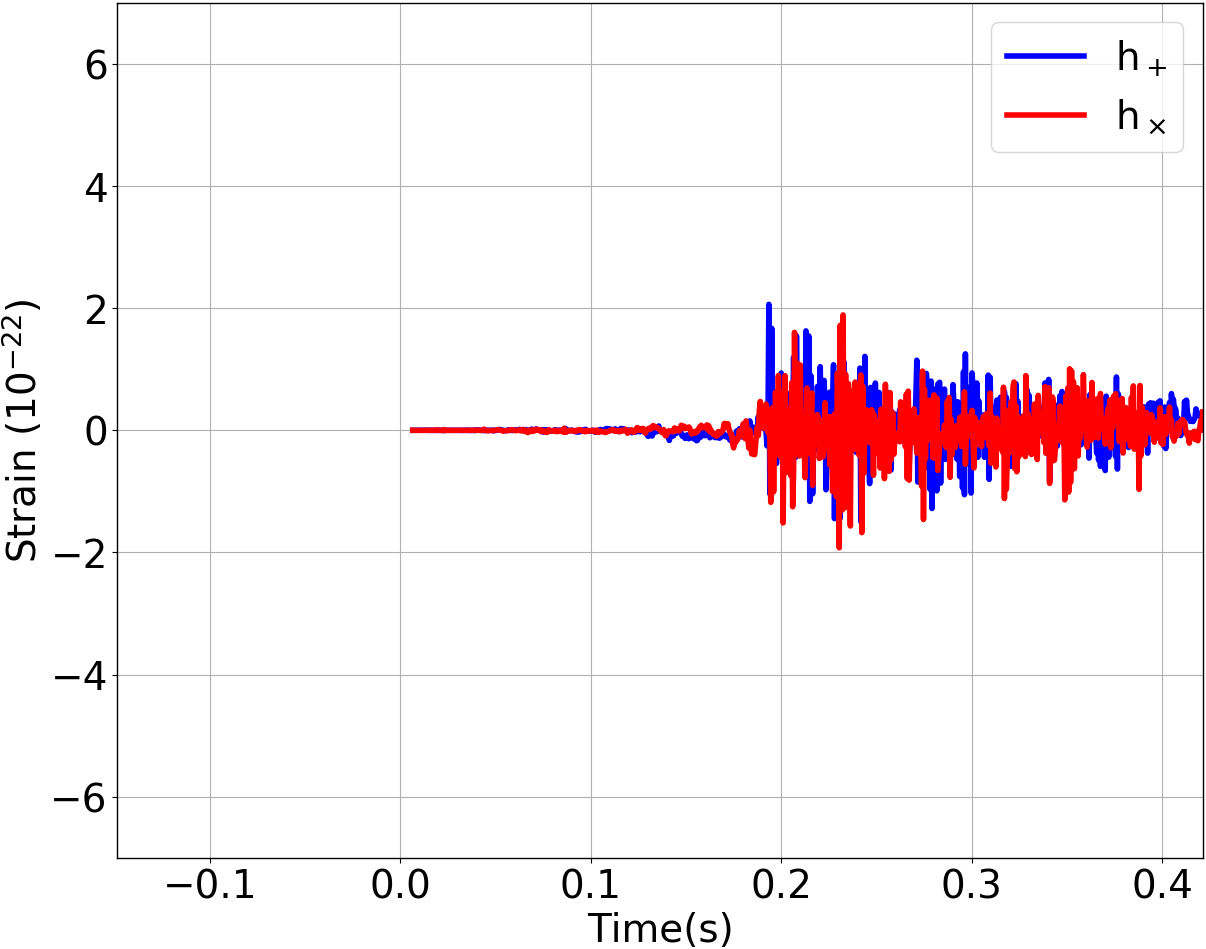}
        }
    \end{center}
    \caption{The waveforms excluded from the training session of the \ac{CNN}
and used as an extra test for the performance of the trained \ac{CNN}. The top 
panels show the waveforms $\text{R3E1AC}$(left) and $\text{R4E1FC\_L}$(right), 
modelled by the magnetorotational mechanism.
The inset plots show the $h_{\times}$ polarisations of the waveform, which are at least one order of magnitude weaker than the 
$h_{+}$ polarisations.
The waveforms shown in the bottom panels are $\text{SFHx}$(left) and $\text{s}20$(right), 
modelled by the neutrino-driven mechanism. 
Shown in the x-axes are the time after core bounce.\label{fig:Extratestwaveform}} 
\end{figure*}

%
%
The previous section proves that using a \ac{CNN}, it is possible to detect and
classify \ac{CCSN} waveforms if appropriate waveforms have been used to train
the \ac{CNN}. However, in reality, it is likely that the \acp{GW} from a
\ac{CCSN} may only be partially similar to the simulated waveforms, while
having some other features that are different or even unexpected. A \ac{CNN}
that is only capable of recognising waveforms with which it is familiar may
prove to be not entirely applicable. Therefore, we apply our trained
\ac{CNN} to waveforms from other studies that were not used during the
training. Similar to the procedure used in~\cite{roma2019astrophysics}, we take
the waveforms $\text{R3E1AC}$ and $\text{R4E1FC\_L}$
from~\cite{scheidegger2010influence} as the test waveforms for the
magnetorotational mechanism. The test waveforms for the neutrino-driven
mechanisms are $\text{s}20$ from~\cite{andresen2017gravitational} and
$\text{SFHx}$ from~\cite{kuroda2016new}. These waveforms are shown
in Fig.~\ref{fig:Extratestwaveform}. The test waveforms in this section should
not be confused with the testing samples in Sec.~\ref{sec:result}, which are
randomly drawn from the same distribution as the training samples.

%
%
\begin{figure}
     \begin{center}
        \subfigure[]{
            \label{fig:extra_waveform_test_neu}
            \includegraphics[width=0.475\textwidth]{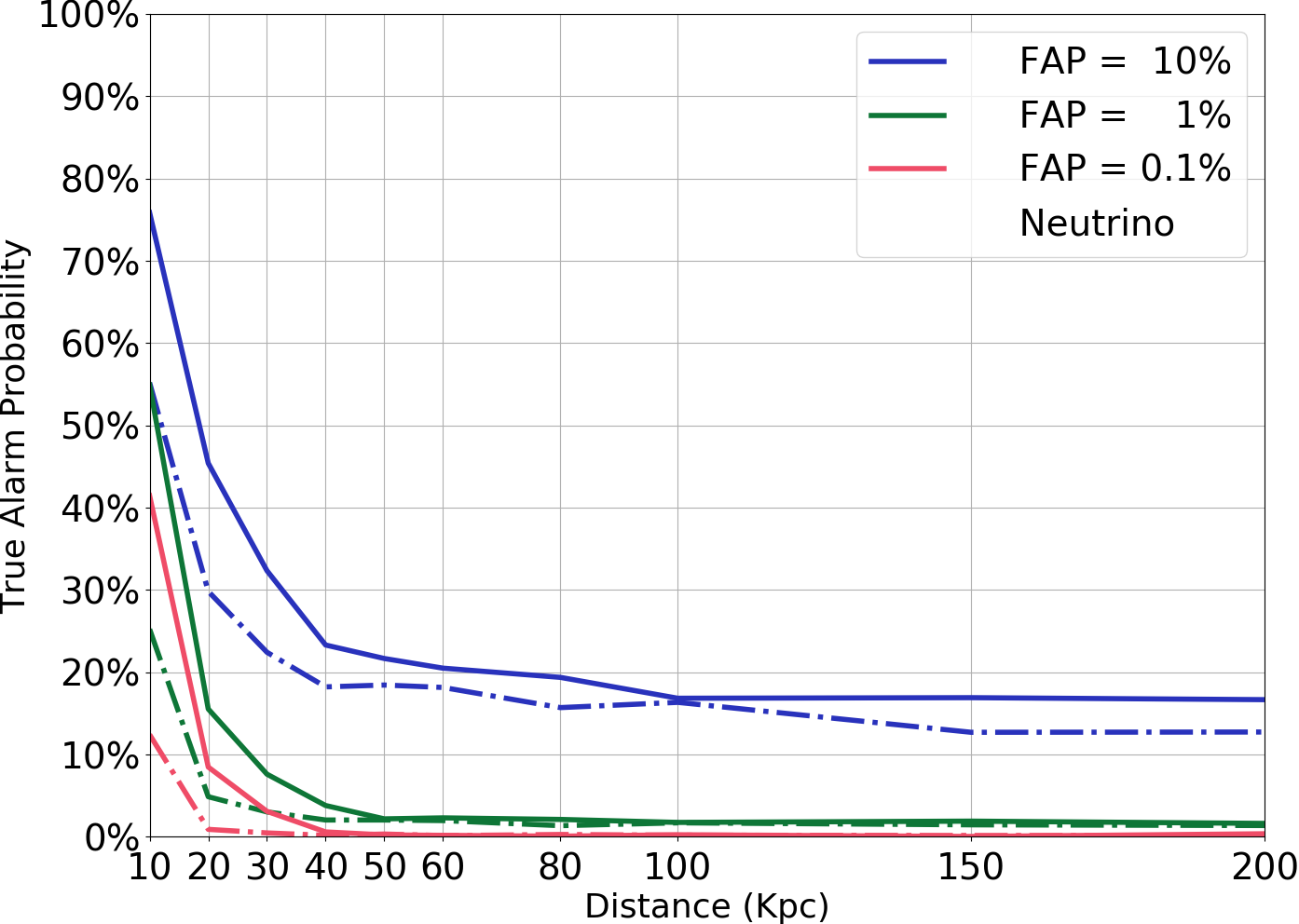}
        }\quad
        \subfigure[]{
            \label{fig:extra_waveform_test_mag}
            \includegraphics[width=0.475\textwidth]{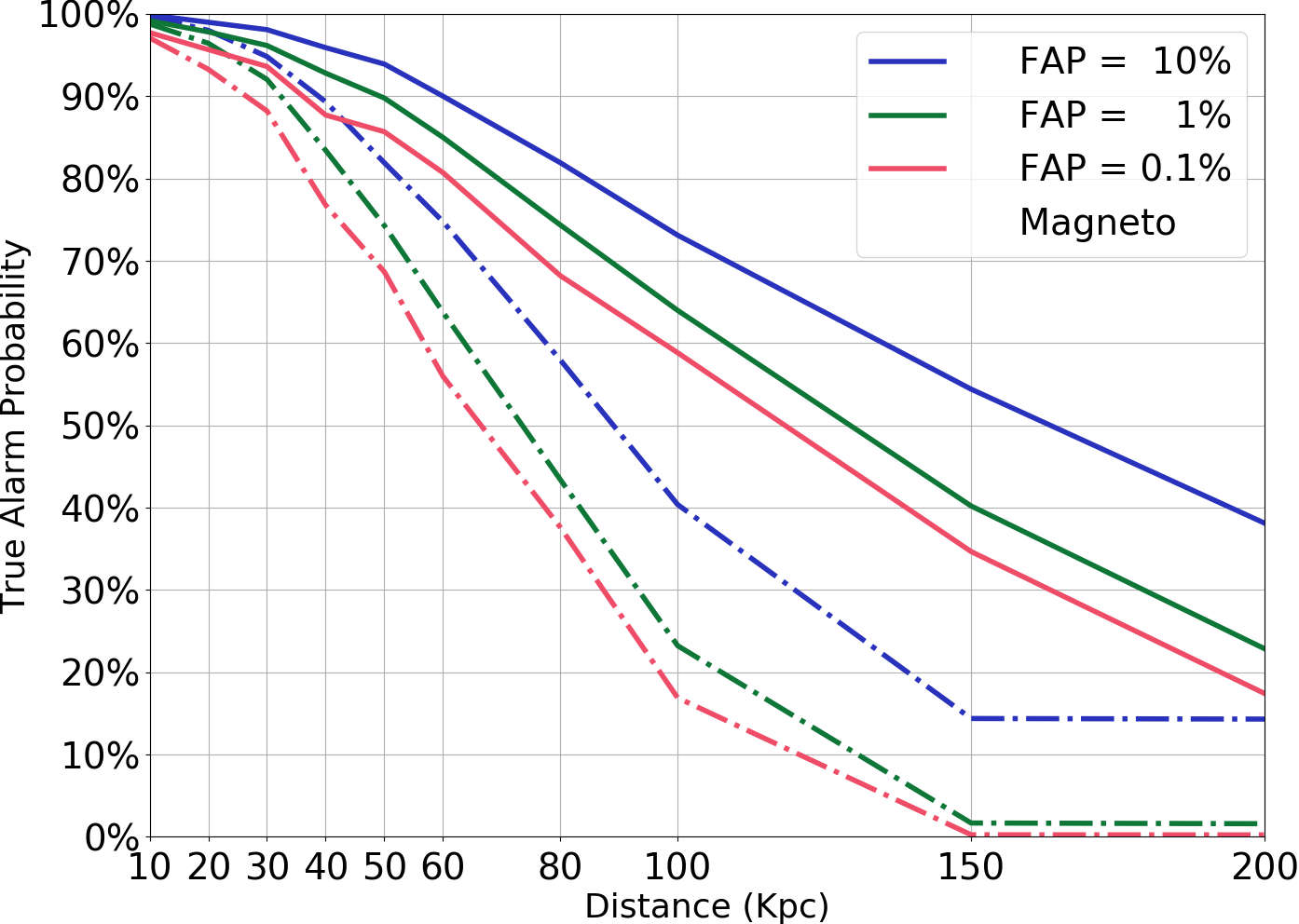}
        }
    \end{center}
    \caption{Efficiency curves showing the ability of the \ac{CNN} in
distinguishing input data.  The upper panel shows the results for the the
waveforms $\text{SFHx}$ and $\text{s}20$ (neutrino-driven mechanism), and the
lower panel shows that for the waveforms $\text{R3E1AC}$ and $\text{R4E1FC\_L}$
(the magnetorotational mechanism).  In both panels, the solid lines show the
\acp{TAP} and \acp{FAP} for H+L+VK, and the dashed lines show those for HLVK.
The waveforms have not been used for the training of the \ac{CNN}.
\label{fig:Extratest}}
\end{figure}

%
%
Using the procedure described in Sec.~\ref{sec:spwf}, we generate $1200$
samples for each distance and therefore each class has $400$
samples. The performance of the \ac{CNN} on the test waveforms is shown in
Fig.~\ref{fig:Extratest}. For waveforms $\text{R3E1AC}$ and $\text{R4E1FC\_L}$
at $50$ kpc and a \ac{FAP} of $10\%$, the \ac{CNN} achieves a \ac{TAP} of $87\%$
with H+L+VK and $59\%$ with HLVK respectively. At $60$ kpc, the \acp{TAP} drops
slightly to $83\%$ and $52\%$ respectively. This indicates that with the
\ac{CNN}, it is possible to detect these waveforms from sources that are at
distances consistent with the Large and Small Magellanic Clouds. For waveforms
$\text{s}20$ and $\text{SFHx}$, the \acp{TAP} are $93\%$ and $70\%$ for the two
networks if the sources are at $10$ kpc and the \ac{FAP} is $10\%$.  The
efficiency for the waveforms $\text{s}20$ and $\text{SFHx}$ is noticeably
higher than that for the neutrino-driven mechanism in Fig.~\ref{fig:eff} for
the same distance. This is because as mentioned before, the detection and
classification efficiency in Fig.~\ref{fig:eff} are the results averaged over
the testing samples. It is possible that individual waveforms may be easier for
the \ac{CNN} to detect. 
\section{Conclusion}\label{sec:conclusion}
%
%
We have demonstrated that a \ac{CNN} can be applied for the purpose of
distinguishing \ac{GW} detector time-series among pure background noise and
\ac{CCSN} explosion mechanisms. We have trained the \ac{CNN} using
$1.8\times10^{5}$ samples of simulated time series for each distance at a
number of distances from $10$ kpc to $200$ kpc. The data samples for each
class consisted of approximately $4\times10^5$ samples. 

%
%
We have shown that with a \ac{GW} detector network of HLVK, when the \ac{FAP}
was $10\%$, once trained, a \ac{CNN} could achieve a \ac{TAP} close to $80\%$
for magnetorotational signals from sources at $60$ kpc. Using a network of
H+L+VK, we showed that the \ac{TAP} is increased to $91\%$.  Both the Large and
Small Magellanic Clouds are within this distance. If the distance is extended
to $150$ or $200$ kpc, a \ac{TAP} of $54\%$ or $38\%$ respectively are still
achievable for H+L+VK, indicating the small possibility of detections within
such distances. 

%
%
For the neutrino-driven mechanism, the weaker amplitudes of the waveforms
result in lower \acp{TAP} at the same distances.  For sources at $10$ kpc, the
trained \ac{CNN}, with a \ac{FAP} of $10\%$, achieved a \ac{TAP} of $55\%$ and
$76\%$ for HLVK and H+L+VK respectively. This indicates a Galactic \ac{CCSN}
event would likely be detectable.

%
%
We used four waveforms that were not used for the training of the \ac{CNN} to
test the performance of the \ac{CNN} in a more realistic situation. We found
that for waveforms $\text{R3E1AC}$ and $\text{R4E1FC\_L}$ from sources at
$60$ kpc, the \acp{TAP} are $83\%$ and $52\%$ for H+L+VK and HLVK respectively.
For waveforms $\text{s}20$ and $\text{SFHx}$, the \acp{TAP} are $93\%$ and
$70\%$ for the two networks respectively. All at \ac{FAP} of $10\%$. 
The results prove the possibility of
using a \ac{CNN} as a tool for the detection and classification of \ac{CCSN}
\ac{GW} signals.
\\
\\
\\
\\
\\
\\
\\
\\
\\
\\
\\
\\
\\
\\
\\
\\
\section{ACKNOWLEDGEMENTS}
We would like to thank Jade Powell and Marek Szczepanczyk for their 
constructive comments on the paper.
We would also like to thank that for the computational resources.
I.S.H. and C.M. are supported by the Science and Technology Research Council
(grant No.~ST/L000946/1) and the European Cooperation in Science and
Technology (COST) action CA17137. 


\end{document}